\documentclass[aps,prb,showpacs,superscriptaddress, preprint numbers,twocolumn]{revtex4-1}
\usepackage{rotating}
\usepackage{epstopdf}
\usepackage{color}
\usepackage{amsmath}
\usepackage{graphicx}
\usepackage{hyperref}
\usepackage{color}
\usepackage{amssymb}
\usepackage{dcolumn}
\usepackage{bm}
\usepackage{epsfig}
\usepackage{array}
\usepackage{subfigure}
\usepackage{upgreek}
\usepackage{wasysym}
\usepackage{slashed}
\usepackage[section]{placeins}
\usepackage{appendix}
\usepackage{subfigure}
\usepackage{adjustbox}
\newcommand{\be}{\begin{equation}} 
\newcommand{\ee}{\end{equation}} 
\newcommand{\bea}{\begin{eqnarray}} 
\newcommand{\eea}{\end{eqnarray}} 
\newcommand{\bqa}{\begin{eqnarray}}
\newcommand{\eqa}{\end{eqnarray}}

\newcommand{\w}{\omega}

\newcommand{\figimpurityA}{
\begin{figure}[htb]
\centering
\hspace{0cm}
\subfigure[noonleline][]
{\label{fig: ac_impurity_freq_whole}\includegraphics[height=35mm,width=40mm]{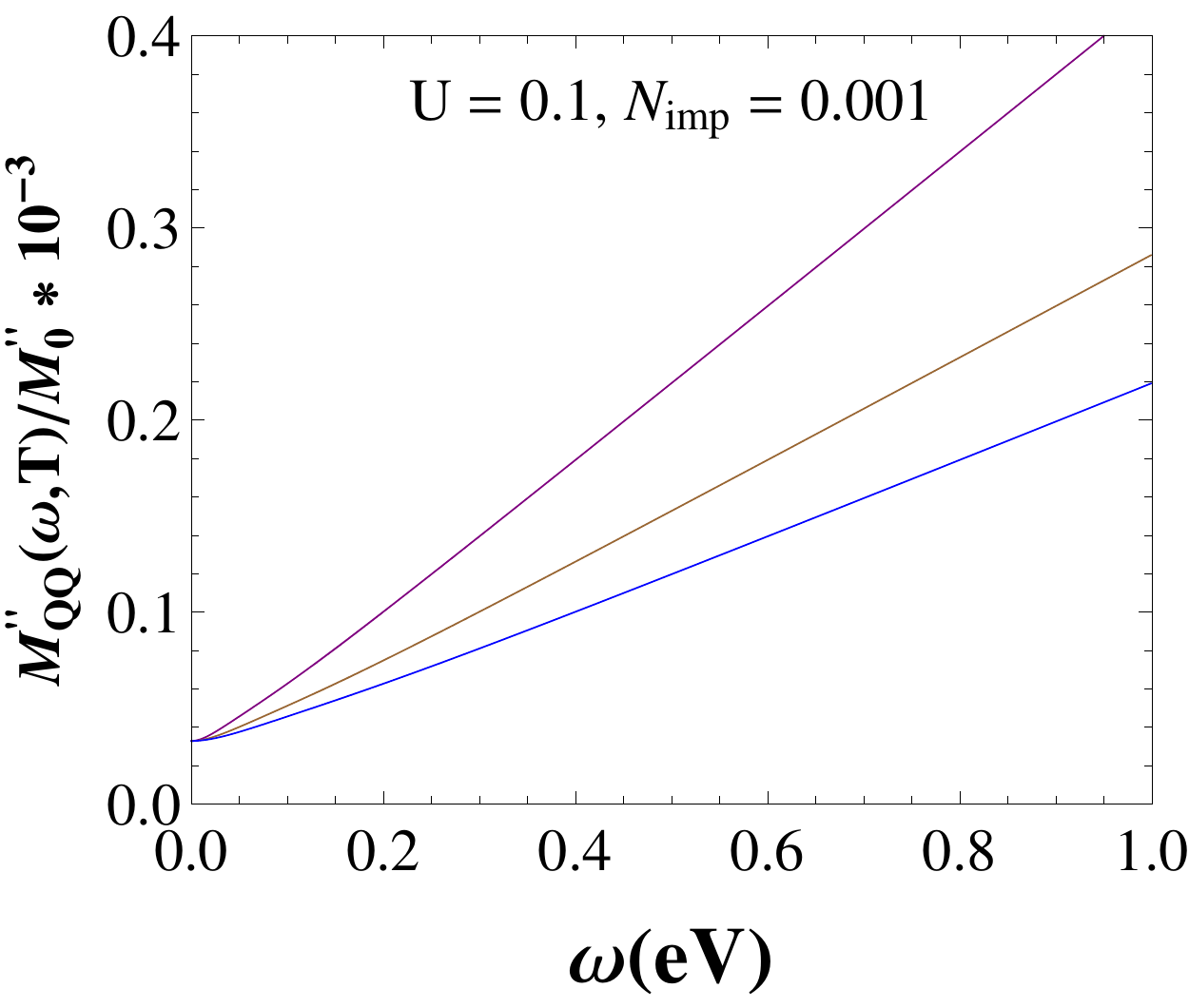}}
\hspace{0cm}
\subfigure[noonleline][]
{\label{fig: ac_impurity_freq_whole_small}\includegraphics[height=35mm,width=40mm]{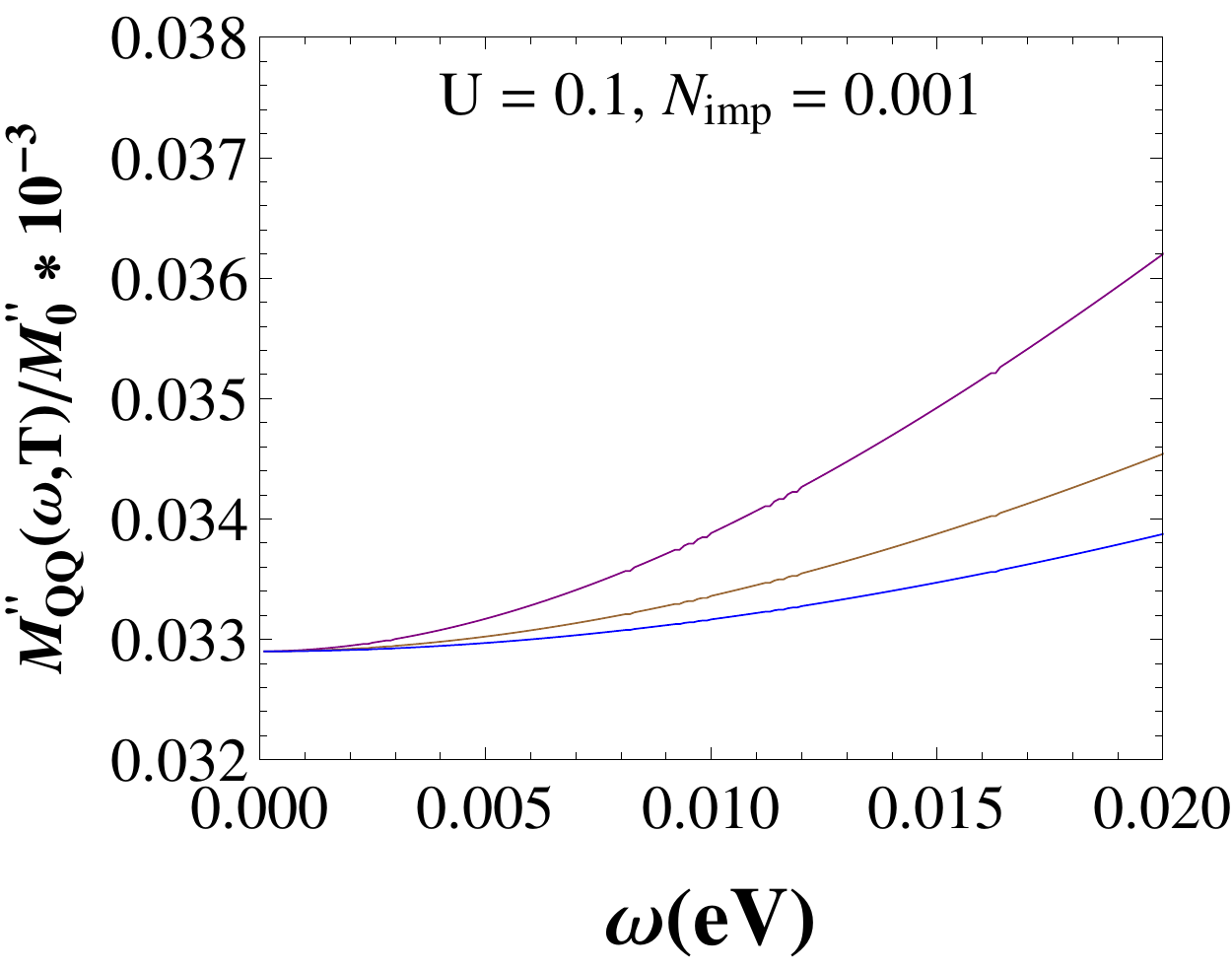}}
\caption{(a): The imaginary part of the thermal memory function for the case of electron-impurity interaction is plotted with frequency at different temperatures such as $200$ (purple), $300$ (brown) and $400$K (blue) at fixed interaction strength $U$ and impurity concentration $N_{\text{imp}}$. (b): The low frequency regime of Fig. \ref{fig: ac_impurity_freq_whole} is elaborated.}
\label{fig: impurity_ac_memory}
\end{figure}
}
\newcommand{\figimpurityC}{
\begin{figure}[htb]
\centering
\hspace{0cm}
\subfigure[noonleline][]
{\label{fig: dc_phonon_temp}\includegraphics[height=35mm,width=40mm]{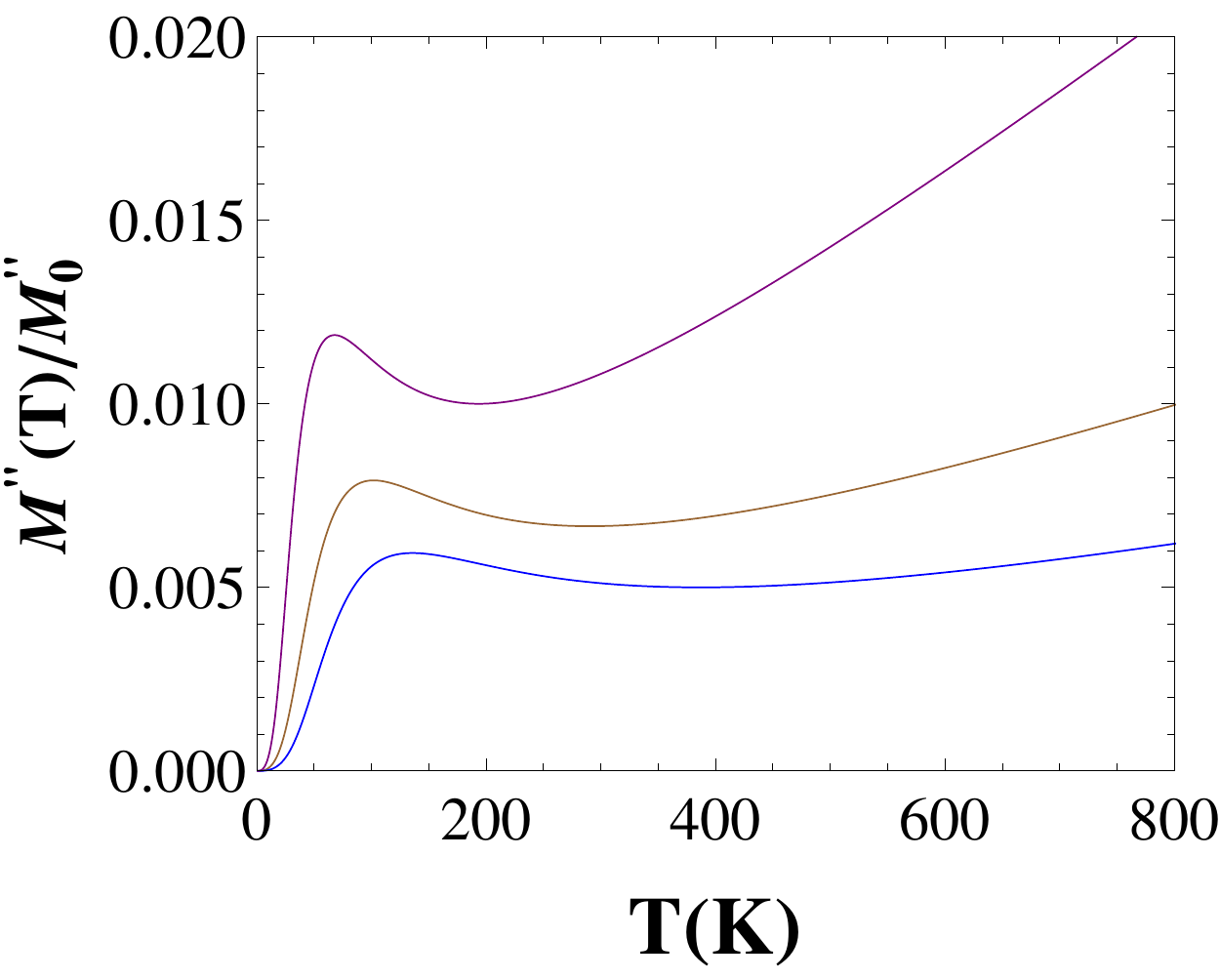}}
\hspace{0cm}
\subfigure[noonleline][]
{\label{fig: dc_thermal_phonon}\includegraphics[height=35mm,width=40mm]{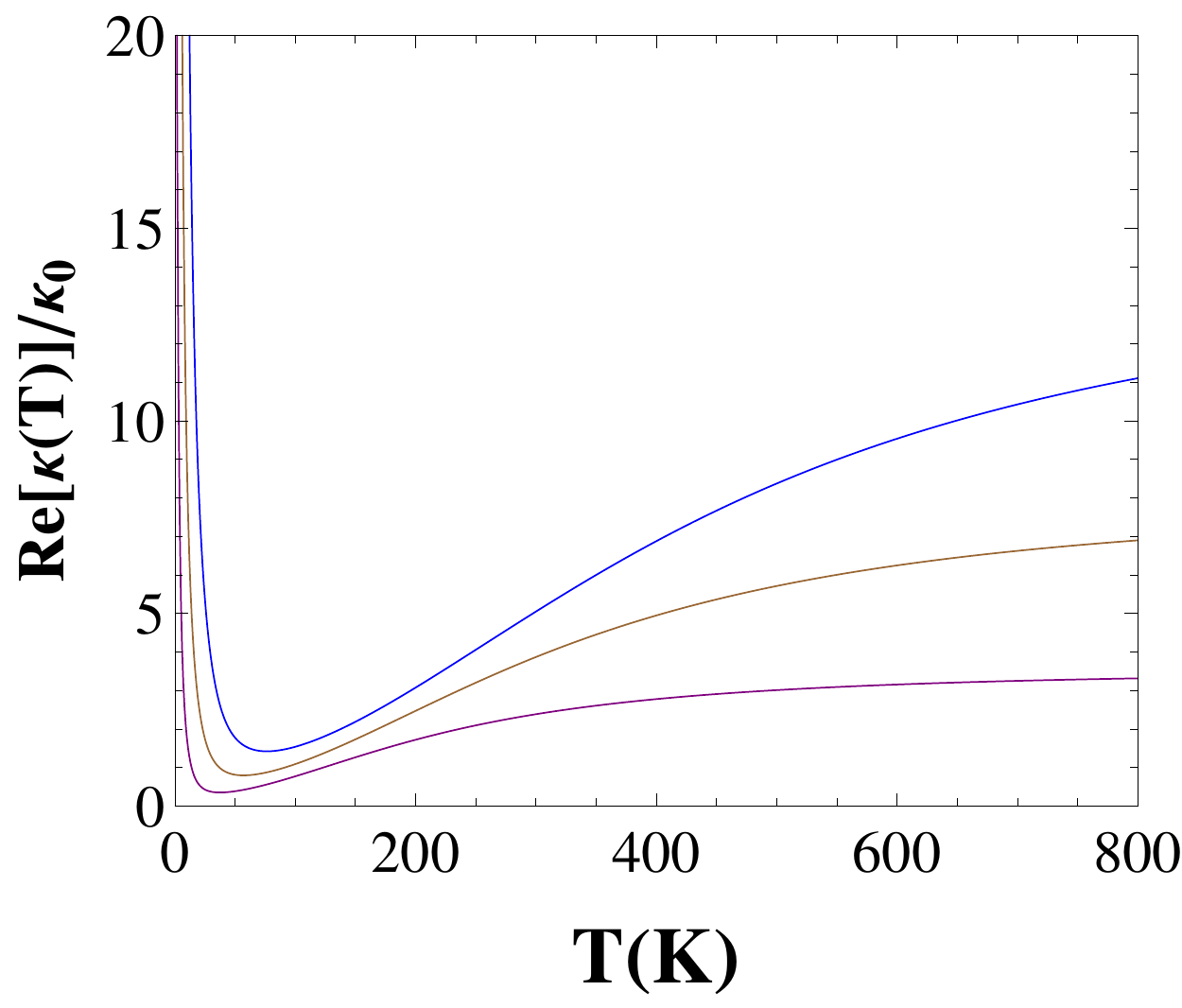}}
\caption{(a): Plot of temperature dependent normalized dc imaginary part of the thermal memory function for electron-phonon interaction at different Debye temperatures such as $200$ (purple), $300$ (brown) and $400$K (blue). (b): The variation of the normalized thermal conductivity with $T$ at same Debye temperatures.}
\label{fig: phonon_dc_thermal_cases}
\end{figure}
}
\newcommand{\figimpurityE}{
\begin{figure}[htb]
\centering
\hspace{0cm}
\subfigure[noonleline][]
{\label{fig: ac_phonon_freq_whole}\includegraphics[height=35mm,width=40mm]{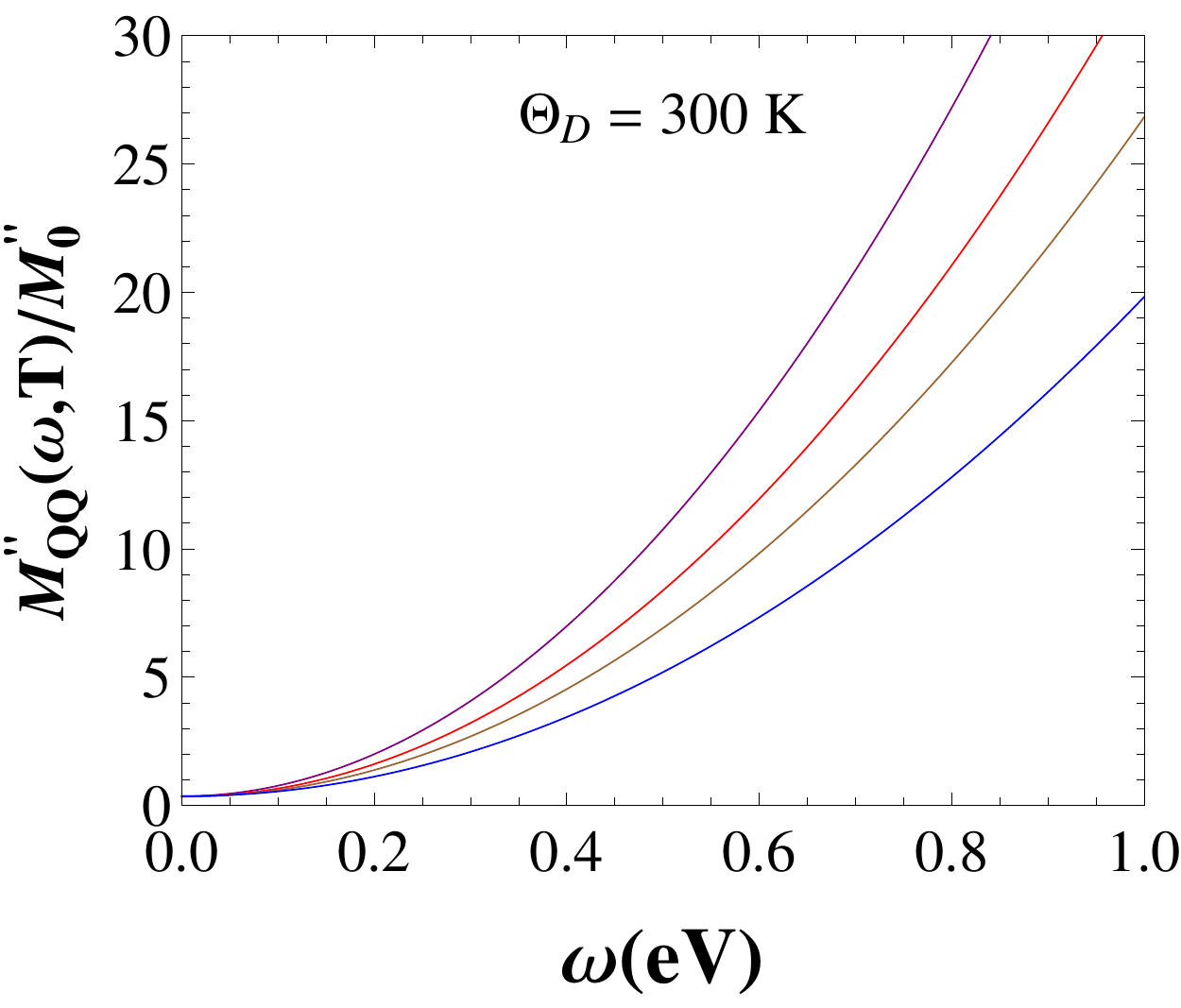}}
\hspace{0cm}
\subfigure[noonleline][]
{\label{fig: ac_phonon_freq_whole_small}\includegraphics[height=35mm,width=40mm]{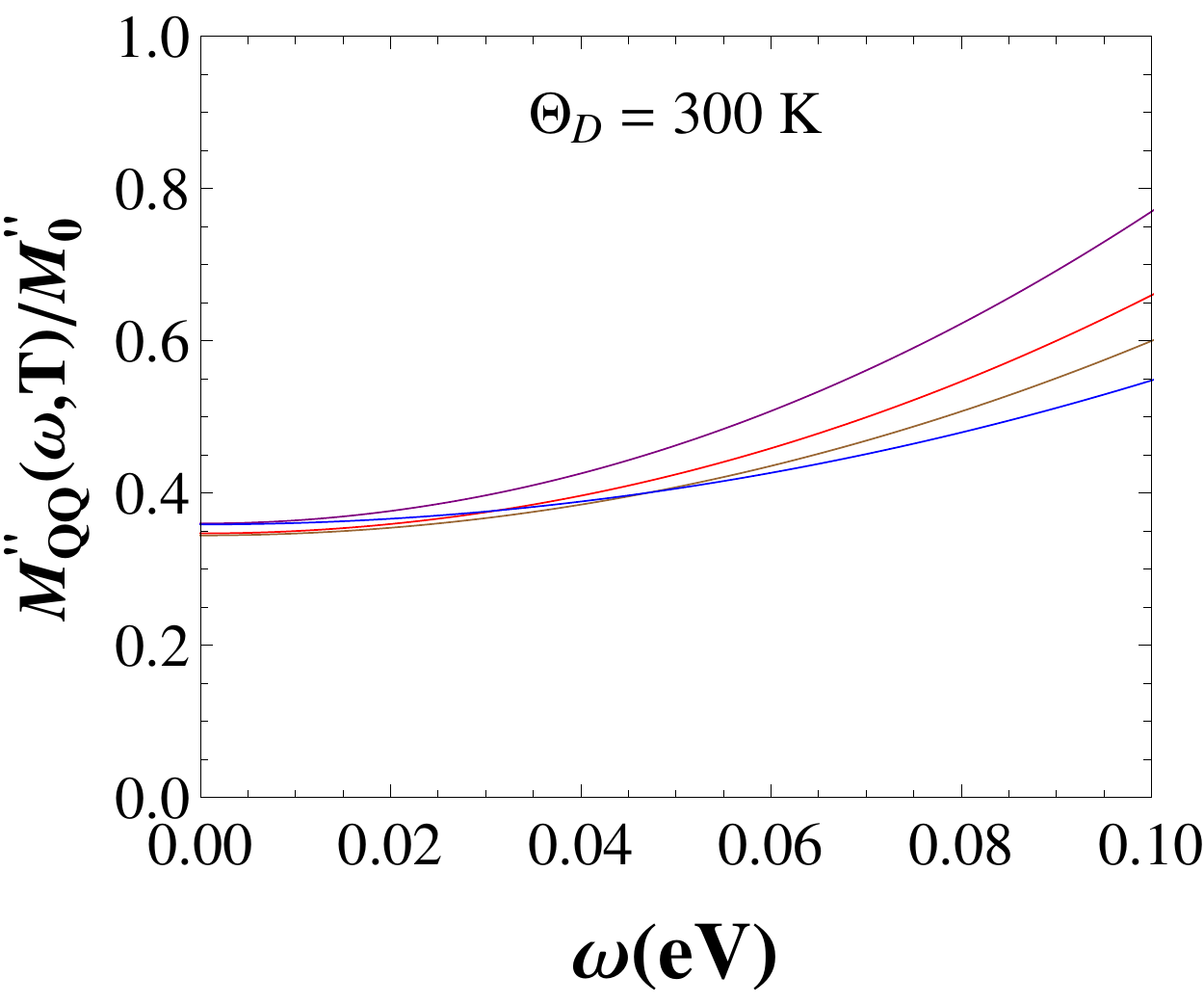}}
\caption{(a): The imaginary part of the thermal memory function for electron-phonon interaction is plotted with frequency at different temperatures such as $200$ (purple), $250$ (red), $300$ (brown) and $400$K (blue) at fixed Debye temperature $\Theta_{D} = 300$K. (b): The low frequency regime of Fig. \ref{fig: ac_phonon_freq_whole} is elaborated.}
\label{fig: phonon_ac_memory}
\end{figure}
}
\newcommand{\figimpurityG}{
\begin{figure}[htb]
\centering
\hspace{0cm}
\subfigure[noonleline][]
{\label{fig: thermal_ac_phonon_freq_whole}
\includegraphics[height=35mm,width=40mm]{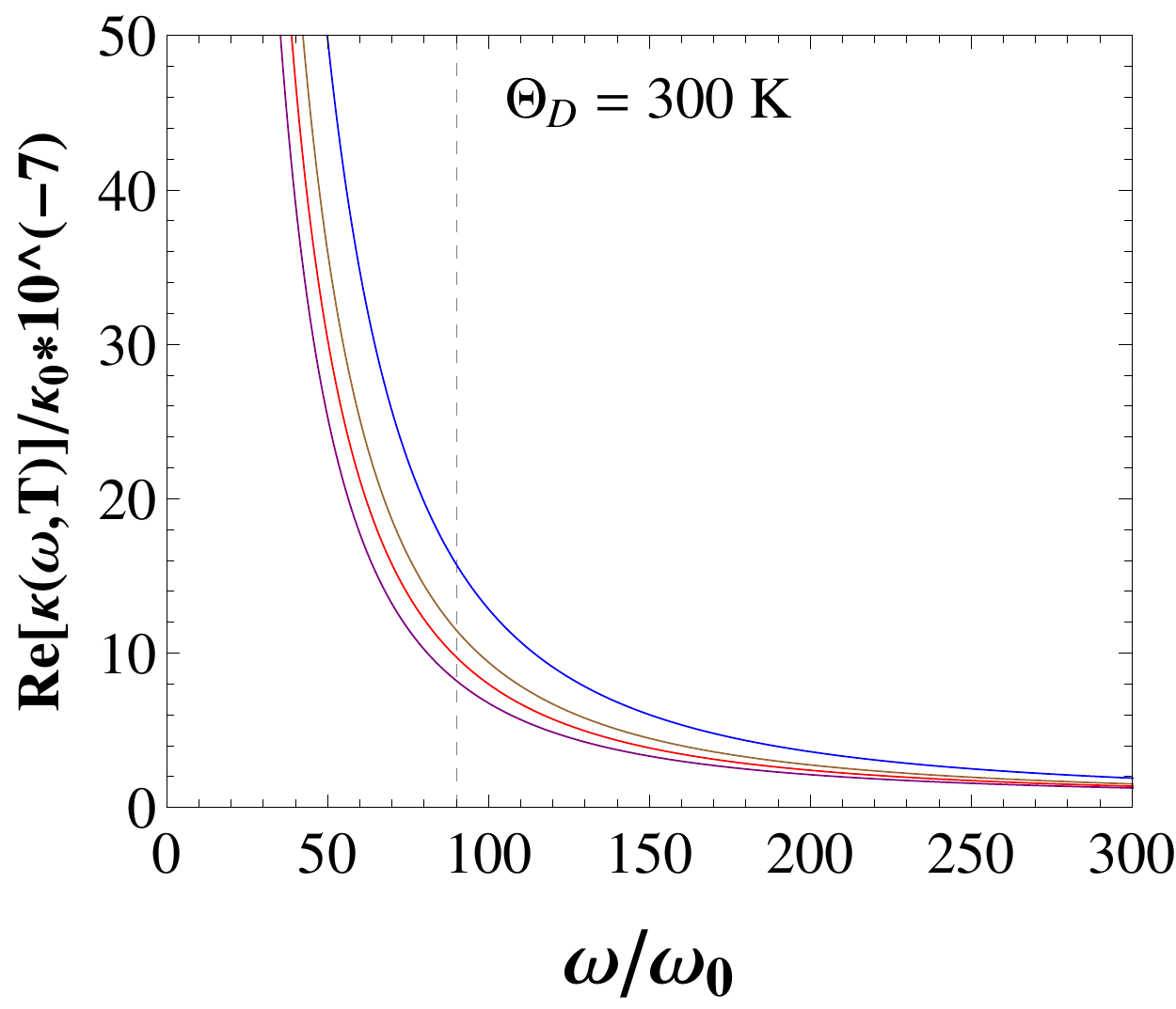}}
\hspace{0cm}
\subfigure[noonleline][]
{\label{fig: thermal_ac_phonon_freq_small}\includegraphics[height=35mm,width=40mm]{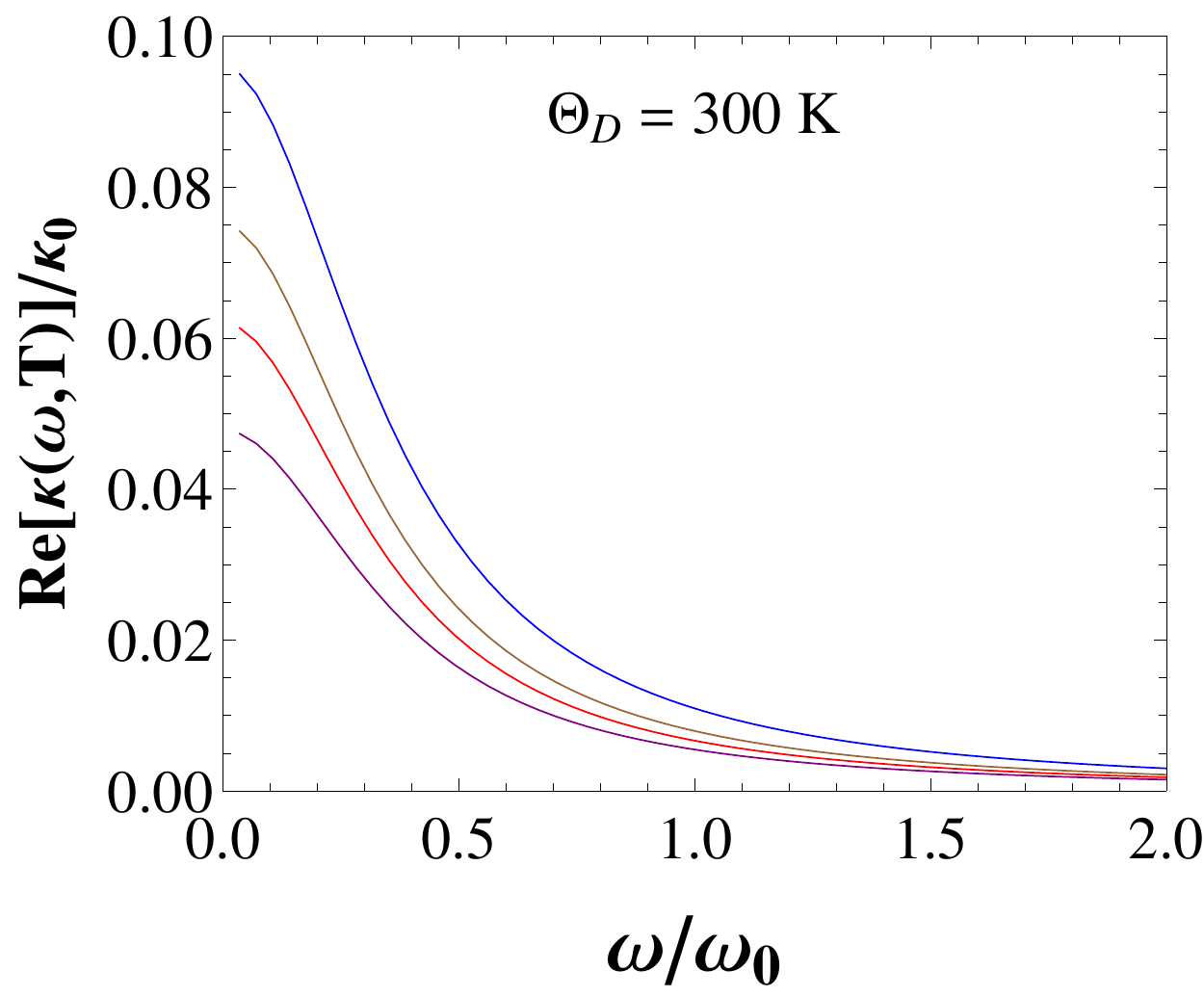}}
\caption{(a). The normalized frequency dependent thermal conductivity is plotted with the ratio $\omega/\omega_{0}$ for electron-phonon interaction at different temperatures such as $200$ (purple), $250$ (red), $300$ (brown) and $400$K (blue) and at Debye temperature $\Theta_{D} = 300$K. Here $\omega_{0}$ is a constant having dimensions of energy and the dashed line corresponds to the scale for Debye frequency cutoff i.e. $\omega_{D}/\omega_{0}$. (b). The low frequency regime of Fig. \ref{fig: thermal_ac_phonon_freq_whole} is elaborated.}
\label{fig: thermal_ac_phonon_freq}
\end{figure}
}

\begin{document}
\title{Theory of the Dynamical Thermal Conductivity of Metals}

\author{Pankaj Bhalla}
\email{pankajbhalla66@gmail.com}
\affiliation{Physical Research Laboratory, Navrangpura, Ahmedabad-380009 India.}
\affiliation{Indian Institute of Technology Gandhinagar-382424, India.}
\author{Pradeep Kumar}
\affiliation{Physical Research Laboratory, Navrangpura, Ahmedabad-380009 India.}
\affiliation{Indian Institute of Technology Gandhinagar-382424, India.}
\author{Nabyendu Das}
\affiliation{Physical Research Laboratory, Navrangpura, Ahmedabad-380009 India.}
\author{Navinder Singh}
\affiliation{Physical Research Laboratory, Navrangpura, Ahmedabad-380009 India.}
\date{\today}
\begin{abstract}
The Mori's projection method, known as memory function method is an important theoretical formalism to study various transport coefficients. In the present work, we calculate the dynamical thermal conductivity in the case of metals using the memory function formalism. We introduce thermal memory functions for the first time and discuss the behavior of thermal conductivity in both zero frequency limit and in the case of non-zero frequencies. We compare our results for the zero frequency case with the results obtained by the Bloch-Boltzmann kinetic approach and find that both approaches agree with each other.
Motivated by some recent experimental advancements, we obtain several new results for the ac or the dynamical thermal conductivity.
\end{abstract}
\date{\today}
\maketitle

\section{Introduction}
\label{sec: intro}
There have been significant advancements in the study of the thermal transport coefficients for complex systems\cite{bridges_02, bhalla_14, chernyshev_15, jain_16, romano_16}.  In such systems, the transport coefficients can be understood via the transport lifetime which captures the role of different interactions such as electron-impurity, electron-phonon and electron-electron interactions. Several methods\cite{bidwell_40, deo_65, wilson_book} based on the Kubo formalism and the Bloch-Boltzmann method have been applied to compute the effects of such interactions on various transport coefficients such as thermal conductivity. The commonly used method is the Bloch-Boltzmann transport method\cite{ziman_book}. Within this approach, it is found that the thermal conductivity $\kappa(T)$ is proportional to the temperature $T$ both in high and low temperature regimes in the case of impurity interactions. While in the case of electron-phonon interactions, it varies as $T^{-2}$ in the low temperature limit ($T << \Theta_{D}$, where $\Theta_{D}$ is the Debye temperature) and saturates to a constant value in the high temperature limit ($T>>\Theta_{D}$)\cite{ziman_book}. These signatures are predicted long ago and are well verified. However, the notion of frequency dependent (dynamical) thermal conductivity was not previously known and hence was not addressed in theoretical discussions.

Recently, the notion of the dynamical thermal conductivity is introduced by Volz et.al\cite{volz_01}. With this idea, the recent experiments access frequency in which $\omega$ dependence cannot be ignored. There it is introduced in the context of its usefulness for the thermal design of microsystems and nanosystems which operates at several GHz clock frequency. Cooling of the Joule heating in such systems is an important issue\cite{volz_01} and it requires detailed understanding of the frequency dependence of the thermal conductivity. In reference[\cite{volz_01}], the dynamical thermal conductivity is introduced in the context of phonon mediated thermal transport in Si crystals. However, in the case of metals, and particularly at certain frequency, the electronic contributions to the thermal conductivity may predominates. We consider that scenario and present the paper to a careful theoretical analysis of the frequency dependent electronic thermal conductivity of metals in various regimes of interest. In a recent computer simulation using molecular dynamics technique, it is found that the phononic thermal conductivity reduces its magnitude at high frequencies\cite{volz_01}. Experimentally, it is also studied in the context of semiconductor alloys and it is found that the magnitude of the phononic thermal conductivity reduces as the frequency increases\cite{koh_07}.

Theoretically, the electronic and the phononic dynamical thermal conductivity is discussed in the recent past by Shastry\cite{shastry_06} and others\cite{shastry_09, dhar_11, ezzahri_12, yang_15} in different contexts such as in open systems, strongly correlated systems, semiconductor crystals, etc. In the present work, we \textit{explicitly} derive the various expressions for the electronic thermal conductivity in case of metal with electron-impurity and electron-phonon interaction.

We use the memory function formalism which was introduced by Mori and Zwanzig\cite{zwanzig_61, zwanzig_61a, mori_65}. It is formulated in several renditions. The commonly used version named projection operator formalism is the most fascinating regarding the physical aspects of the systematic approximations. The main motivation of this approach is the determination of the time correlation function in quantum or classical many body systems in a systematic way\cite{forster_95, fulde_12, pires_88, berne_66, berne_70, harp_70, arfi_92, maldague_77, nabyendu_15, nabyendu_16}.

We calculate for the first time, \textit{the dynamical thermal memory functions} for the case of electron-impurity and electron-phonon interactions. It is directly related to the dynamical thermal conductivity viz. $\kappa(z,T) \sim \frac{1}{z+M_{QQ}(z,T)}$, where $M_{QQ}(z,T)$ is the thermal memory function and $z$ is the complex frequency. The details of $M_{QQ}(z,T)$ will be discussed in the next section. The results in the zero frequency limit are consistent with the results predicted using Bloch-Boltzmann approach. We also calculate the dynamical thermal memory functions in different frequency regimes and discuss the effects of the impurity and the phonon scattering on it.

This paper is organized as follows: we review the basics of the memory function formalism in Sec.\ref{sec: memory}. Later in Sec.\ref{sec: thermal conductivity}, we introduce the model Hamiltonian and then calculate the thermal memory functions for the case of electron-impurity and electron-phonon interactions. Then, we discuss its behavior in different frequency and temperature regimes. Here we also calculate the asymptotic behavior of the thermal conductivity in the presence of these interactions. The results for the zero frequency case is compared with the results previously obtained by the Boltzmann approach and we find good agreement. We make several predictions in frequency dependence cases in Sec.\ref{sec: results}. Finally, in Sec.\ref{sec: conclusion}, we conclude. 

\section{Memory Function Formalism}
\label{sec: memory}
Before embarking into the detailed calculation of the thermal memory function, let us first briefly review the general framework of the memory function formalism in this section.

Consider two operators $A$ and $B$ corresponding to two different physical observables. Their correlation function is defined as\cite{kubo_57, zubarev_60, kadanoff_63}
\be
\chi_{AB}(t)=\langle A(t) ; B(0) \rangle,
\ee
where $\langle \cdots \rangle $ denotes the thermal average and $t$ is the time variable. The Laplace transform of the correlation function in the complex frequency domain can be expressed as
\be
\chi_{AB}(z) = \langle \langle A ; B \rangle \rangle_{z} = -i\int_{0}^{\infty} e^{izt} \langle [A(t), B] \rangle dt.
\label{eqn: definitioncorrelation}
\ee
Here $[\cdot,\cdot]$ represents the commutator between two operators, $z$ is the complex frequency variable and the outer angular bracket $\langle \cdots \rangle $ in $\langle \langle A ; B \rangle \rangle_{z}$ refers to the Laplace transform.\\
In frequency space, the equation of motion of this correlation function can be cast in the following form
\be
z\langle \langle A ; B \rangle \rangle_{z} = \langle [A,B] \rangle + \langle \langle [A,H]; B \rangle \rangle_{z}.
\label{eqn: eqmcorrelation}
\ee
Here $H$ represents the total Hamiltonian of the system. In the present work, we are interested in calculating the thermal current-thermal current correlation function. Thus we replace both the general operators $A$ and $B$ by the thermal current operator $J_{Q}$ and the equation (\ref{eqn: eqmcorrelation}) takes the form
\be
z\langle \langle J_{Q} ; J_{Q} \rangle \rangle_{z} = \langle [J_{Q},J_{Q}] \rangle + \langle \langle [J_{Q},H]; J_{Q} \rangle \rangle_{z}.
\ee
Here the first term in the right hand side contains equal time commutator $[J_{Q},J_{Q}]$ which identically vanishes. Thus, $z\langle \langle J_{Q} \vert J_{Q} \rangle \rangle_{z} = \langle \langle [J_{Q},H]; J_{Q} \rangle \rangle_{z}$. Again applying equation of motion on $\langle \langle [J_{Q},H]; J_{Q} \rangle \rangle_{z}$, one obtains 
\bea \nonumber
&& z\langle \langle J_{Q} ; J_{Q} \rangle \rangle_{z} \\ && = \frac{\langle \langle [J_{Q},H]; [J_{Q},H] \rangle \rangle_{z=0} - \langle \langle [J_{Q},H]; [J_{Q},H] \rangle \rangle_{z}}{z}.
\eea
Finally, the correlation function can be expressed as
\bea \nonumber && 
\chi_{QQ}(z,T) \\ &&=\frac{\langle \langle [J_{Q},H]; [J_{Q},H] \rangle \rangle_{z=0} - \langle \langle [J_{Q},H]; [J_{Q},H] \rangle \rangle_{z}}{z^2}.
\label{eqn: correlation}
\eea
Following the Ref[\cite{gotze_72, navinder_16}], the correlation function $\chi_{QQ}(z,T)$ and the memory function $M_{QQ}(z)$ are related as
\be
M_{QQ}(z,T)=z\frac{\chi_{QQ}(z,T)}{\chi_{QQ}^{0}(T)-\chi_{QQ}(z,T)},
\ee
where $\chi_{QQ}^{0}(T)$ is the static thermal current-thermal current correlation function. This above expression is identical to that in the case of electrical transport.\\
On considering the assumption that $\chi_{QQ}(z,T)/\chi_{QQ}^{0}(T)$ is smaller than one, the above expression with the leading order term can be expressed as
\be
M_{QQ}(z,T)\approx z\frac{\chi_{QQ}(z,T)}{\chi_{QQ}^{0}(T)}.
\label{eqn: memoryassumption}
\ee
The validity of this approximation is discussed in detail in the references\cite{bhalla_16, bhalla_16a} for the electrical transport and same should follow to the case of thermal transport.\\
Using Eq.(\ref{eqn: correlation}) and (\ref{eqn: memoryassumption}), the thermal memory function can be written as
\bea \nonumber &&
M_{QQ}(z,T)\\ &&=\frac{\langle \langle [J_{Q},H]; [J_{Q},H] \rangle \rangle_{z=0} - \langle \langle [J_{Q},H]; [J_{Q},H] \rangle \rangle_{z}}{z\chi_{QQ}^{0}(T)}. 
\label{eqn: memory}
\eea
This is an expression for the complex thermal memory function in terms of the thermal force-thermal force correlation. Further the thermal conductivity can be written in terms of the thermal memory function as follows,
\be
\kappa(z,T) = i\frac{1}{T}  \frac{\chi_{QQ}^{0}(T)}{z+M_{QQ}(z,T)}.
\label{eqn: thermal}
\ee
This is a general expression for the thermal conductivity in a memory function formalism (proof is given in Appendix \ref{app: thermal_relation_memory}). Here $M_{QQ}(z,T)$ is the thermal memory function which provides the information about the effects of various interactions such as electron-impurity and electron-phonon interactions on the thermal conductivity $\kappa(z, T)$. The specific cases are discussed in detail in the next section.

\section{Thermal Conductivity}
\label{sec: thermal conductivity}
\subsection{Model Hamiltonian}
In this work, we consider a system in which electrons interact with impurities and phonons. The total Hamiltonian of such a system takes the form
\be
H = H_{0} + H_{\text{imp}} + H_{\text{ep}} + H_{\text{ph}}.
\label{eqn: hamiltonian}
\ee
Here the first term in the right hand side of the above Eq. corresponds to the unperturbed part which is expressed as
\be
H_{0} = \sum_{\textbf{k} \sigma} \epsilon_{\textbf{k}} c^{\dagger}_{\textbf{k} \sigma} c_{\textbf{k} \sigma},
\ee
where $\epsilon_{\textbf{k}}$ is the energy dispersion for free electrons, $c_{\textbf{k} \sigma}$ and $c^{\dagger}_{\textbf{k} \sigma}$ are annihilation and creation operators having crystal momentum $\textbf{k}$ and spin $\sigma$. The second term is the perturbed Hamiltonian for the electron-impurity interactions which is described as
\be
H_{\text{imp}} = N^{-1} \sum_{i} \sum_{\textbf{k} \textbf{k}' \sigma} \langle \textbf{k} \vert U^{i} \vert \textbf{k}^{\prime} \rangle c^{\dagger}_{\textbf{k} \sigma} c_{\textbf{k}' \sigma}.
\label{eqn: impurityhamiltonian}
\ee
Here $N$ represents the number of lattice cells, $U^{i}$ refers for impurity interaction strength and sum over $i$ index refers for the number of impurity sites. Here the unit cell volume is taken as unity. The third term of Eq.(\ref{eqn: hamiltonian}) describes the interacting Hamiltonian for electron-phonon interactions which is defined as
\be
H_{\text{ep}} = \sum_{\textbf{k} \textbf{k}' \sigma} \left[ D(\textbf{k}-\textbf{k}') c^{\dagger}_{\textbf{k} \sigma} c_{\textbf{k}' \sigma} b_{\textbf{k}-\textbf{k}'} + H.c. \right].
\label{phononhamiltonian}
\ee
Here $b_{\textbf{q}}$($b_{\textbf{q}}^{\dagger}$) is the phonon annihilation(creation) operator having momentum $\textbf{q}$. The electron-phonon matrix element $D(\textbf{q})$ can be considered in the following form\cite{ziman_book}
\be
D(\textbf{q}) = \frac{1}{\sqrt{2m_{i}N\omega_{q}}} q C(q),
\label{eqn: matrixelement}
\ee
where $m_{i}$ is the ion mass, $\omega_{q}$ is the phonon dispersion. $C(q)$ is a slowly varying function of the phonon momentum which in case of metal is considered as $1/\rho_{F}$, where $\rho_{F}$ is the density of the states (DOS) at the Fermi surface\cite{ziman_book}. The last term of the Hamiltonian represents free phonons and is given by
\be
H_{\text{ph}} = \sum_{q} \omega_{q} \left( b_{q}^{\dagger} b_{q} +\frac{1}{2} \right).
\label{freephononhamiltonian}
\ee
With this Hamiltonian, we proceed to the calculation of the thermal memory functions.
\subsection{Thermal Memory functions}
\label{sec: thermalmemoryfunction}
To compute the thermal memory functions, we need to define the heat current\cite{mahan_book} which is the energy current where energy is measured with respect to the electronic chemical potential $\mu$. In an operator form, it can be written as
\bea
J_{Q} = \frac{1}{m} \sum_{\textbf{k}} \textbf{k}.\hat{n} (\epsilon_{\textbf{k}} - \mu) c_{\textbf{k}}^{\dagger} c_{\textbf{k}},
\label{eqn: thermalcurrent}
\eea
where $\hat{n}$ is the unit vector parallel to the direction of heat current and $m$ is the electron mass.\\
Using this definition, let us focus on the calculation of the thermal memory function and hence thermal conductivity. In general, the $M_{QQ}(z,T)$ is a complex valued function of frequency having both real and imaginary parts. Its imaginary part describes the scattering rate due to the presence of different interactions such as electron-impurity and electron-phonon interactions. On the other hand, the real part describes mass enhancement.
\subsubsection{Electron-Impurity Interaction}
\label{sec: electron-impurity}
In the presence of only electron-impurity interactions, the thermal memory function defined in Eq.(\ref{eqn: memory}) is computed by considering the total Hamiltonian $H = H_{0} + H_{\text{imp}}$.

To compute it, we first evaluate the commutator of $J_{Q}$ and $H$. Since $J_{Q}$ commutes with free part of Hamiltonian, $H_{0}$, then $[J_{Q},H] = [J_{Q},H_{\text{imp}}]$. Thus using the Eq.(\ref{eqn: impurityhamiltonian}) and (\ref{eqn: thermalcurrent}), the commutator becomes
\bea \nonumber
[J_{Q},H] &=& \frac{1}{m N} \sum_{i} \sum_{\textbf{k} \textbf{k}' \sigma}  \langle \textbf{k} \vert U^{i} \vert \textbf{k}' \rangle \\ 
&& \left(\textbf{k} (\epsilon_{\textbf{k}} - \mu) - \textbf{k}'(\epsilon_{\textbf{k}'}- \mu) \right).\hat{n}c^{\dagger}_{\textbf{k} \sigma} c_{\textbf{k}' \sigma}.
\label{eqn: thermalimpuritycommutator}
\eea
Using the above expression, the Laplace transform and the thermal average of the inner product $\langle \langle [J_{Q},H] ; [J_{Q},H]\rangle \rangle_{z}$ becomes
\bea \nonumber
= \frac{1}{m^2 N^2} \sum_{i j} \sum_{\textbf{k} \textbf{k}' \sigma} \sum_{\textbf{p} \textbf{p}' \tau} \langle \textbf{k} \vert U^{i} \vert \textbf{k}' \rangle \langle \textbf{p} \vert U^{j} \vert \textbf{p}' \rangle \\ \nonumber
\left(\textbf{k} (\epsilon_{\textbf{k}} - \mu) - \textbf{k}' (\epsilon_{\textbf{k}'} - \mu) \right).\hat{n} \\ \nonumber
\left(\textbf{p} (\epsilon_{\textbf{p}}- \mu) - \textbf{p}' (\epsilon_{\textbf{p}'} - \mu) \right).\hat{n} \\
 \langle \langle c^{\dagger}_{\textbf{k} \sigma} c_{\textbf{k}' \sigma} ; c^{\dagger}_{\textbf{p} \tau} c_{\textbf{p}' \tau} \rangle \rangle_{z}.
\label{eqn: doubleannularbracket}
\eea
By considering the case of dilute impurity i.e. $i=j$ and performing the ensemble average using Eq.(\ref{eqn: definitioncorrelation}) followed by integration over time, the Eq.(\ref{eqn: doubleannularbracket}) takes the following form
\bea \nonumber
&=& \frac{2 N_{\text{imp}}}{m^2 N^2} \sum_{\textbf{k} \textbf{k}'}  \vert \langle \textbf{k} \vert U \vert \textbf{k}' \rangle \vert^{2} \left[\left(\textbf{k} (\epsilon_{\textbf{k}} - \mu) - \textbf{k}' (\epsilon_{\textbf{k}'} - \mu) \right).\hat{n}\right]^{2} \\ 
&& \times \frac{f_{\textbf{k}}-f_{\textbf{k}'}}{z+\epsilon_{\textbf{k}}-\epsilon_{\textbf{k}'}}.
\eea
Here $N_{\textnormal{imp}}$ represents the impurity concentration, the factor $2$ is due to the electronic spin degeneracy and $f_{\textbf{k}} = \frac{1}{e^{\beta (\epsilon_{\textbf{k}} - \mu)}+1}$ is the Fermi distribution function and $\beta$ is the inverse of the temperature.\\
Substituting the above Eq. in Eq.(\ref{eqn: memory}) and on performing the analytic continuation $z \rightarrow \omega + i \eta$, $\eta \rightarrow 0^+$, the imaginary part of the thermal memory function becomes
\bea \nonumber
M''_{QQ}(\omega, T) &=& \frac{2 \pi}{N^{2}} \frac{N_{\text{imp}}}{\chi_{QQ}^{0}(T) m^2} \sum_{\textbf{k} \textbf{k}'}  \vert \langle \textbf{k} \vert U \vert \textbf{k}' \rangle \vert^{2} \\ \nonumber
&& \times \left[ \left(\textbf{k} (\epsilon_{\textbf{k}} -\mu) - \textbf{k}' (\epsilon_{\textbf{k}'}-\mu) \right).\hat{n} \right]^{2} \\ 
&& \times \frac{f_{\textbf{k}}-f_{\textbf{k}'}}{\omega} \delta(\omega +\epsilon_{\textbf{k}}-\epsilon_{\textbf{k}'}).
\label{eqn: memoryimpurity2}
\eea
To reduce the Eq. further, it is convenient to assume that the system has cubic symmetry. Then on averaging over all directions, we obtain
\bea \nonumber &&
\left[ \left(\textbf{k} (\epsilon_{\textbf{k}} -\mu) - \textbf{k}' (\epsilon_{\textbf{k}'}-\mu) \right).\hat{n} \right]^{2} \\
&=& \frac{1}{3}\vert \textbf{k} (\epsilon_{\textbf{k}} -\mu) - \textbf{k}' (\epsilon_{\textbf{k}'}-\mu) \vert^2.
\eea
Using the above Eq. along with the assumption that $U$ is independent of momentum, the Eq.(\ref{eqn: memoryimpurity2}) can be written in the integral form
\bea \nonumber
M''_{QQ}(\omega,T)&=& \frac{U^2 N_{\textnormal{imp}}}{3 (2\pi)^5 m^2 \chi_{QQ}^{0}(T)} \int \frac{d\epsilon_{\textbf{k}}}{v_{\textbf{k}}} k^{2} \sin\theta d\theta d\phi \\
\nonumber && \int \frac{d\epsilon_{\textbf{k}'}}{v_{\textbf{k}'}}  k'^{2} \sin\theta' d\theta' d\phi' \\ \nonumber
&& \vert \textbf{k} (\epsilon_{\textbf{k}} -\mu) - \textbf{k}' (\epsilon_{\textbf{k}'}-\mu) \vert^{2} \\
&& \frac{f_{\textbf{k}}-f_{\textbf{k}'}}{\omega} \delta(\omega +\epsilon_{\textbf{k}}-\epsilon_{\textbf{k}'}).
\eea
For our convenience, we drop the subscript $\textbf{k}$ from all $\epsilon_{\textbf{k}}$ in further calculations and solve one of the energy integral using the property of delta function. In a typical metal, the Fermi energy is very large (is of the order of $10^4$K). On the other hand the experiments are usually performed at temperature of the order of $10^2$K. Thus, electrons from a small region of width $k_{B}T$ (in the present case $k_{B} =1$) around the Fermi surface participate in the scattering events. Hence, we assume that the magnitudes of $\textbf{k}$ and $\textbf{k}'$ are equal to $k_{F}$, the Fermi wave vector. Thus, the imaginary part of the thermal memory function takes the following form 
\bea \nonumber
M''_{QQ}(\omega, T) &=& \frac{N_{\textnormal{imp}} U^2 k_{F}^{4}}{6 \pi^3 \chi_{QQ}^{0}(T)} \int d\epsilon \left((\epsilon -\mu)^2 +  (\epsilon -\mu + \omega)^2 \right) \\
&& \times \frac{f(\epsilon - \mu)-f(\epsilon - \mu + \omega)}{\omega}.
\label{eqn: memoryimpurity4}
\eea
Substituting $\frac{\epsilon -\mu}{T} = \eta$ and $\frac{\omega}{T} = x$, the above expression can be written in simpler form as
\bea \nonumber
M''_{QQ}(\omega, T) &=& \frac{N_{\textnormal{imp}} U^2 k_{F}^{4} T^{2}}{6 \pi^3  \chi_{QQ}^{0}(T)} \int_{0}^{\infty} d\eta \frac{\eta^2 + (\eta + x)^2}{x} \\
&& \left[ \frac{1}{e^{\eta} +1} - \frac{1}{e^{\eta + x} +1} \right].
\label{eqn: memoryimpurity3}
\eea
This is the final expression for the imaginary part of the thermal memory function due to the impurity interactions. Here we assume that the electronic kinetic energy is higher than the temperature $T$. Further in various frequency and temperature limits, its behavior can be discussed as follows:\\
\textbf{Case-I: In the dc limit i.e. $\omega \rightarrow 0$}\\
In this limit, the Eq.(\ref{eqn: memoryimpurity3}) reduces to
\bea
M''_{QQ}(T) &=& \frac{N_{\text{imp}}}{3 \pi^3} \frac{U^2k_{F}^{4} T^2}{\chi_{QQ}^{0}(T)} \int_{0}^{\infty} d\eta \eta^2 \frac{e^{\eta}}{(e^{\eta} + 1)^2}.
\label{eqn: memoryimpurity5}
\eea
This concludes that the temperature dependent imaginary part of the thermal memory function, also known as thermal scattering rate, $1/\tau_{th}$ varies with temperature as $T^2/ \chi_{QQ}^{0}(T)$. Since the static correlation function $\chi_{QQ}^{0}(T)$ is directly proportional to the square of temperature (proof is given in Appendix \ref{app: static_derivation}). Thus, $1/\tau_{th}$ in the zero frequency limit is independent of the temperature. This result agrees with the Bloch-Boltzmann result. On the other hand, due to the symmetry relations of the thermal memory function\cite{gotze_72}, its real part becomes identically zero in the dc limit. On substituting this in the expression for the thermal conductivity (Eq.\ref{eqn: thermal}), we find that the real part of the thermal conductivity depends on the temperature as
\be
\text{Re}[\kappa(T)] = \frac{1}{T} \frac{\chi_{QQ}^{0}(T)}{M''_{QQ}(T)}.
\ee
Using Eq.(\ref{eqn: memoryimpurity5}) and (\ref{eqn: static_expression}) (mentioned in the Appendix \ref{app: static_derivation}), the above Eq. for the thermal conductivity reduces to
\bea \nonumber
\text{Re}[\kappa(T)] &=& \frac{1}{72} \frac{\pi k_{F}^{2}}{N_{\text{imp}}U^{2}m^{2}} T \\
\text{i.e.} \hspace*{2mm} \text{Re}[\kappa(T)] &\propto& T.
\eea
This result is in accord with the result predicted earlier using Boltzmann's equation approach (Eq.(\ref{eqn: thermal_boltzmann_impurity}) in Appendix \ref{app: boltzmann_impurity}). \\
\textbf{Case-II: In the finite frequency limit}\\
In the high frequency limit i.e. $\omega >> T$
, the imaginary part of the thermal memory function becomes
\bea \nonumber
M''_{QQ}(\omega, T) &\approx& \frac{N_{\textnormal{imp}}U^2 k_{F}^{4} T^{2}}{6 \pi^3 \chi_{QQ}^{0}(T)} \int_{0}^{\infty} d\eta x  \left[ \frac{1}{e^{\eta} +1} - \frac{1}{e^{\eta + x} +1} \right].\\
&\approx& \frac{N_{\textnormal{imp}}U^2 k_{F}^{4} T^{2}}{6 \pi^3 \chi_{QQ}^{0}(T)} \int_{0}^{\infty} d\eta \frac{1}{e^{\eta} +1} \frac{\omega}{T}
\label{eqn:dcthermalmemory}
\eea
This yields that the thermal memory function or the thermal scattering rate approximately varies linearly with the frequency and inversely with the temperature. While in the opposite case $\omega << T$, the leading order term in the Eq.(\ref{eqn:dcthermalmemory}) becomes 
\bea \nonumber
M''_{QQ}(T) &\approx& \frac{N_{\textnormal{imp}}U^2 k_{F}^{4} T^{2}}{6 \pi^3 \chi_{QQ}^{0}(T)} \int_{0}^{\infty} d\eta \frac{\eta^2}{e^{\eta} +1}  \left(2-\frac{\omega}{T} \right).\\
\eea
These results are summarized in the table(\ref{tab:impuritytable}).
\begin{table}[b]
\caption{The thermal scattering rate due to the electron-impurity interaction in different frequency and temperature domains.}
\begin{center}
\begin{adjustbox}{max width = 8.6cm}
 \begin{tabular}{|c| c|} 
 \hline
 $\omega = 0$ & $\omega \ne 0$\\ 
 \hline
 $1/\tau_{\text{th}} \sim T^{0}$ & \begin{tabular}{c} 
 \begin{tabular}{c| c} 
 $\omega >> T$ & $\omega << T$\\ 
 \hline
 \begin{tabular}{c} 
 $1/\tau_{\text{th}} \sim \frac{\omega}{T}$ \\ [1ex] 
 \end{tabular} & \begin{tabular}{c} 
 $1/\tau_{\text{th}} \sim \left(2-\frac{\omega}{T}\right) $ \\  
 \end{tabular}
  \end{tabular} \\ 
\end{tabular} \\ [1ex] 
 \hline 
  \end{tabular}
  \end{adjustbox}
  \end{center}
  \label{tab:impuritytable}
 \end{table}

\subsubsection{Electron-Phonon Interaction}
\label{sec: electron-phonon}
Now consider that the system has only electron-phonon interaction. Then, the thermal memory function can be calculated in a similar fashion as is done in the case of the impurity interaction. Here the total Hamiltonian is considered as $H = H_{0} + H_{\text{ep}} + H_{\text{ph}}$. The thermal current commutes with the free electron and the free phonon parts of the Hamiltonian. Thus, we are left with the commutator of the thermal current $J_{Q}$ and the interaction term $H_{\text{ep}}$ which is expressed as
\bea \nonumber
[J_{Q}, H_{\text{ep}}] &=& \frac{1}{m} \sum_{\textbf{k} \textbf{k}' \sigma} \left(\textbf{k} (\epsilon_{\textbf{k}} - \mu) - \textbf{k}' (\epsilon_{\textbf{k}'} - \mu) \right).\hat{n} \\
&& \left( D(\textbf{k}-\textbf{k}') c^{\dagger}_{\textbf{k} \sigma} c_{\textbf{k}' \sigma} b_{\textbf{k}-\textbf{k}'} - H.c. \right).
\label{thermalphononcommutator}
\eea
Using the above commutation relation, $\langle \langle [J_{Q},H_{\text{ep}}] ; [J_{Q},H_{\text{ep}}]\rangle \rangle_{z}$ can be cast in the following form
\bea \nonumber
&=& \frac{1}{m^2} \sum_{\textbf{k} \textbf{k}' \sigma} \sum_{\textbf{p} \textbf{p}' \tau} \left( \textbf{k} (\epsilon_{\textbf{k}} - \mu) - \textbf{k}' (\epsilon_{\textbf{k}'} - \mu) \right).\hat{n} \\ \nonumber
&& \left( \textbf{p} (\epsilon_{\textbf{p}} - \mu) - \textbf{p}'(\epsilon_{\textbf{p}'} - \mu) \right).\hat{n} \\ \nonumber
&& \left( D(\textbf{k}-\textbf{k}') D^*(\textbf{p}-\textbf{p}') \langle \langle c^{\dagger}_{\textbf{k} \sigma} c_{\textbf{k}' \sigma} b_{\textbf{k}-\textbf{k}'} ;  c^{\dagger}_{\textbf{p}' \tau} c_{\textbf{p} \tau} b^{\dagger}_{\textbf{p}-\textbf{p}'}\rangle \rangle_{z} \right. \\ \nonumber
&& \left. - D^{*}(\textbf{k}-\textbf{k}') D(\textbf{p}-\textbf{p}') \langle \langle c^{\dagger}_{\textbf{k}' \sigma} c_{\textbf{k} \sigma} b^{\dagger}_{\textbf{k}-\textbf{k}'} ;  c^{\dagger}_{\textbf{p} \tau} c_{\textbf{p}' \tau} b_{\textbf{p}-\textbf{p}'}\rangle \rangle_{z} \right).\\
\eea
On further simplifications, the above expression reduces to 
\bea \nonumber
&=& \frac{2}{m^2} \sum_{\textbf{k} \textbf{k}'} \left[\left( \textbf{k} (\epsilon_{\textbf{k}} - \mu) - \textbf{k}' (\epsilon_{\textbf{k}'} - \mu) \right).\hat{n}\right]^{2} \vert D(\textbf{k}-\textbf{k}') \vert^{2}  \\ \nonumber
&&\left( f_{\textbf{k}}(1-f_{\textbf{k}'})(1+n) -(1-f_{\textbf{k}})f_{\textbf{k}'} n \right) \\ \nonumber
&& \left\lbrace \frac{1}{z+\epsilon_{\textbf{k}}-\epsilon_{\textbf{k}'} - \omega_{\textbf{k}-\textbf{k}'}} - \frac{1}{z+\epsilon_{\textbf{k}'}-\epsilon_{\textbf{k}} + \omega_{\textbf{k}-\textbf{k}'}} \right\rbrace, \\
\eea
where $n = \frac{1}{e^{\beta \omega_{q}}-1}$ is the Boson distribution function at a temperature $1/\beta$.\\
On substituting the above Eq. in the thermal memory function Eq.(\ref{eqn: memory}) and then performing the analytic continuation $z \rightarrow \omega + i\eta$, $\eta \rightarrow 0^{+}$, the imaginary part of the thermal memory function can be written as
\bea \nonumber
M''_{QQ}(\omega, T) &=& \frac{2\pi}{\chi_{QQ}^{0}(T) m^2} \sum_{\textbf{k} \textbf{k}'} \left[\left( \textbf{k} (\epsilon_{\textbf{k}} - \mu) - \textbf{k}'(\epsilon_{\textbf{k}'} - \mu) \right).\hat{n}\right]^{2} \\ \nonumber
&& \vert D(\textbf{k}-\textbf{k}') \vert^{2} (1-f_{\textbf{k}})f_{\textbf{k}'} n \\ \nonumber
&& \left\lbrace \frac{e^{\omega/T}-1}{\omega} \delta(\epsilon_{\textbf{k}}-\epsilon_{\textbf{k}'} - \omega_{\textbf{k}-\textbf{k}'} + \omega) \right. \\
&& \left. + (\text{terms with $\omega \rightarrow -\omega$}) \right\rbrace.
\label{eqn: memoryphonon2}
\eea
To evaluate the above Eq., we use the law of conservation of energy $\epsilon_{\textbf{k}} = \epsilon_{\textbf{k}'} - \omega_{q}$ and conservation of momentum $\textbf{q} = \textbf{k}' - \textbf{k}$ which simplify a factor appearing in the Eq.(\ref{eqn: memoryphonon2}) as follows
\bea \nonumber
\left[(\textbf{k} (\epsilon_{\textbf{k}} - \mu) - \textbf{k}'(\epsilon_{\textbf{k}'} - \mu).\hat{n}\right]^2 &=& \left[(\omega_{q}\textbf{k}' + (\epsilon_{\textbf{k}} - \mu)\textbf{q}).\hat{n}\right]^2. \\
\eea
For simplicity, we consider that the system has cubic symmetry as considered in the case of impurity. Then on averaging over all directions, we obtain
\bea\nonumber
\left[(\omega_{q}\textbf{k}' + (\epsilon_{\textbf{k}} - \mu)\textbf{q}).\hat{n}\right]^2 &=& \frac{1}{3} \left\lbrace \omega_{q}^2 k'^2 + q^2 (\epsilon_{\textbf{k}} - \mu)^2 \right. \\
&& \left. + \omega_{q}(\epsilon_{\textbf{k}} - \mu) q^2  \right\rbrace. 
\label{eqn: cubicsymmetry}
\eea
Substituting the Eq.(\ref{eqn: cubicsymmetry}) in (\ref{eqn: memoryphonon2}) and on converting the summations to integrals, we get
\bea \nonumber
M''_{QQ}(\omega, T) &=& \frac{N^2}{3 \chi_{QQ}^{0}(T) m^2 (2\pi)^5}  \int \frac{d\epsilon_{\textbf{k}}}{v_{\textbf{k}}} k^2 \sin\theta  d\theta d\phi \\ \nonumber
&& \int \frac{d\epsilon_{\textbf{k}'}}{v_{\textbf{k}'}} k'^2 \sin\theta' d\theta' d\phi' \int dq \vert D(q) \vert^{2}  \\ \nonumber
&& \delta(q - \vert \textbf{k}-\textbf{k}' \vert) (1-f_{\textbf{k}})f_{\textbf{k}'} n \\ \nonumber
&& \left\lbrace \omega_{q}^2 k'^2 + q^2 (\epsilon_{\textbf{k}} - \mu)^2 + \omega_{q}(\epsilon_{\textbf{k}} - \mu) q^2  \right\rbrace \\ \nonumber
&& \left\lbrace \frac{e^{\omega/T}-1}{\omega} \delta(\epsilon_{\textbf{k}}-\epsilon_{\textbf{k}'} - \omega_{\textbf{k}-\textbf{k}'} + \omega) \right. \\
&& \left. + (\text{terms with $\omega \rightarrow -\omega$}) \right\rbrace.
\eea
Following the argument as quoted in the impurity case, for low energy scattering, we consider the magnitudes of $\textbf{k}$ and $\textbf{k}'$ of the order of $\textbf{k}_{F}$. With these facts and solving one of the energy integrals, the above Eq. reduces to
\bea \nonumber
M''_{QQ}(\omega, T) &=& \frac{N^2}{12 \pi^3} \frac{1}{\chi_{QQ}^{0}(T)} \int_{0}^{\infty} d\eta \int_{0}^{q_{D}} dq q \vert D(q) \vert^{2}     \\ \nonumber
&& \frac{1}{e^{y} - 1} \frac{1}{e^{-\eta}+1} \left\lbrace \omega_{q}^2 k_{F}^2 + q^2 \eta^2 T^2+ \omega_{q} \eta T q^2  \right\rbrace \\ \nonumber
&&  \left[ \frac{1}{e^{\eta-y-x}+1} \frac{e^{x}-1}{x}  \right. \\
&& \left.+ (\text{terms with $\omega \rightarrow -\omega$}) \right].
\eea
Here we introduce new dimensionless variables $\frac{\epsilon_{\textbf{k}} - \mu}{T} = \eta$, $\frac{\omega_{q}}{T} = y$ and $\frac{\omega}{T} = x$.  Now integrating over $\eta$, we obtain 
\bea \nonumber
M''_{QQ}(\omega, T) &=& \frac{N^2 T^6}{12 \pi \chi_{QQ}^{0}(T)} \left(\frac{q_{D}}{\Theta_{D}} \right)^4 \int_{0}^{\Theta_{D}/T} dy y^3 \vert D(y) \vert^{2} \\ \nonumber
&& \left[ \frac{(x-y)}{e^{x-y}-1}  \frac{e^{x}-1}{x(e^{y}-1)} \right. \\ \nonumber
&& \left.\left\lbrace \frac{k_{F}^2 }{\pi^2} \left(\frac{\Theta_{D}}{q_{D} T} \right)^2 + \frac{1}{3} + \frac{(x-y)^2}{\pi^2} \right.\right. \\ \nonumber
&& \left.\left.+ \frac{1}{2\pi^2} y (x-y) \right\rbrace \right. \\
&& \left.+ (\text{terms with $\omega \rightarrow - \omega$}) \right].
\label{eqn: memoryphonon3}
\eea
Substituting the phonon matrix element using the Eq.(\ref{eqn: matrixelement}), the thermal memory function is simplified to
\bea \nonumber
M''_{QQ}(\omega, T) &=& \frac{N}{24 \pi m_{i} \rho_{F}^2} \frac{T^7}{\chi_{QQ}^{0}(T)} \left( \frac{q_{D}}{\Theta_{D}} \right)^6  \int_{0}^{\Theta_{D}/T} dy y^4 \\ \nonumber
&& \left[ \frac{(x-y)}{e^{x-y}-1} \frac{e^{x}-1}{x(e^{y}-1)} \left\lbrace \frac{k_{F}^2}{\pi^2}\left( \frac{\Theta_{D}}{q_{D}T}\right)^2 \right. \right. \\ \nonumber
&& \left. \left. + \frac{1}{3} + \frac{(x-y)^2}{3 \pi^2} + \frac{1}{2 \pi^2} y(x-y) \right\rbrace \right. \\
&& \left.+ (\text{terms with $\omega \rightarrow - \omega$}) \right].
\label{eqn: memoryphonon4}
\eea
This is the frequency and the temperature dependent thermal memory function for the case of electron-phonon interaction. In certain regimes of temperature and frequency, this can be solved analytically and are discussed as follows:\\
\textbf{Case-I: In the dc limit i.e. $\omega \rightarrow 0$}\\
In this limit, the Eq.(\ref{eqn: memoryphonon4}) reduces to
\bea \nonumber
M''_{QQ}(T) &=& \frac{N}{12 \pi m_{i} \rho_{F}^2} \frac{T^7}{\chi_{QQ}^{0}(T)} \left( \frac{q_{D}}{\Theta_{D}}\right)^6 \int_{0}^{\Theta_{D}/T} dy \\ \nonumber
&& \frac{y^5 e^{y}}{(e^{y}-1)^2}  \left\lbrace \frac{k_{F}^2}{\pi^2 T^2} \left(\frac{\Theta_{D}}{q_{D}}\right)^2 + \frac{1}{3} - \frac{1}{6 \pi^2} y^2 \right\rbrace. \\
\label{eqn: memoryphonon5}
\eea
In the high temperature limit i.e. when the temperature is much more than the Debye temperature ($T >> \Theta_{D}$), the second term within the curly brackets contributes more as compared to the other terms. Because the other terms varies inversely as square of the temperature, they contribute less then the second term (i.e. $1/3$). Hence, the thermal memory function $M''_{QQ}(T)$ with leading term can be approximated as
\bea \nonumber
M''_{QQ}(T) & \approx & \frac{N}{36 \pi m_{i} \rho_{F}^2}\frac{T^7}{\chi_{QQ}^{0}(T)} \left(\frac{q_{D}}{\Theta_{D}}\right)^6 \\ \nonumber
&& \int_{0}^{\Theta_{D}/T} dy \frac{y^5 e^{y}}{(e^{y}-1)^2}.
\label{eqn: memoryphonon6}
\eea
\bea \nonumber
M''_{QQ}(T) &=& \frac{N \Theta_{D}^4}{144 \pi m_{i} \rho_{F}^2} \left(\frac{q_{D}}{\Theta_{D}}\right)^6 \frac{T^3}{\chi_{QQ}^{0}(T)}.\\
\label{eqn: memoryphonon8}
\eea
Thus on considering the temperature variation of the static thermal correlation function, we find that the imaginary part of the dc thermal memory function varies linearly with the temperature in the high temperature regime. On substituting this in Eq.(\ref{eqn: thermal}), we find that the real part of the thermal conductivity varies as
\be
\text{Re} [\kappa(T)] = \text{constant}.
\ee
In the low temperature limit i.e. when the temperature is much less than the Debye temperature ($T<< \Theta_{D}$), the first term and the third term in the Eq.(\ref{eqn: memoryphonon5}) contributes more to the thermal memory function as compared to the second term. If we consider $q_{D}$ to be smaller than the $k_{F}$, then the first term dominates over third term. Thus using this fact $M''_{QQ}(T)$ becomes
\bea \nonumber
M''_{QQ}(T) &\approx& \frac{N k_{F}^2}{12 \pi^3 m_{i} \rho_{F}^2} \left(\frac{q_{D}}{\Theta_{D}}\right)^6 \frac{T^5}{\chi_{QQ}^{0}(T)}\\ 
&& \int_{0}^{\infty} dy y^5 \frac{e^{y}}{(e^{y}-1)^2}. 
\label{eqn: memoryphonon9}
\eea
The above Eq. tells that the imaginary part of the thermal memory function or the thermal scattering rate varies as $T^3$ ($1/\tau_{\text{th}} \propto T^{3}$). As argued in the impurity case, the mass renormalization is zero. Thus, we find that the real part of the thermal conductivity (Eq.(\ref{eqn: thermal})) which varies inversely as square of the temperature i.e.
\be
\text{Re}[ \kappa(T)] \propto T^{-2}.
\ee
These results in different temperature regimes are in accord with the results obtained by the Boltzmann equation approach \cite{wilson_book, ziman_book} and with the experimental results\cite{ziman_54, rosenberg_55, klemens_56}. In Appendix \ref{app: boltzmann_phonon}, we compare these results with the results from the Bloch-Boltzmann equation and we observe agreement.\\
\textbf{Case-II: In the finite frequency case}\\
In the high frequency limit i.e. when frequency is much higher than the Debye frequency ($\omega >> \omega_{D}$), the thermal memory function (Eq.(\ref{eqn: memoryphonon4})) becomes
\bea \nonumber
M''_{QQ}(\omega, T) &\approx& \frac{N}{12 \pi m_{i} \rho_{F}^2} \frac{T^7}{\chi_{QQ}^{0}(T)} \left( \frac{q_{D}}{\Theta_{D}} \right)^6  \int_{0}^{\Theta_{D}/T} dy  \\  \nonumber
&& \frac{y^4}{e^{y}-1} \left\lbrace \frac{k_{F}^2}{\pi^2}\left( \frac{\Theta_{D}}{q_{D}T}\right)^2 + \frac{1}{3} + \frac{1}{3 \pi^2} \frac{\omega^2}{ T^2}\right\rbrace. \\
\label{eqn: memoryphononhighfreq}
\eea
In the high temperature limit i.e. $T >> \Theta_{D}$ and $\omega << T$, the second term of Eq.(\ref{eqn: memoryphononhighfreq}) contributes more over the other terms. Thus, the imaginary part of the thermal memory function becomes
\bea \nonumber
M''_{QQ}(\omega, T) &\approx& \frac{N}{36 \pi m_{i} \rho_{F}^2} \left(\frac{q_{D}}{\Theta_{D}}\right)^6 \frac{T^7}{\chi_{QQ}^{0}(T)} \\  
&& \times \int_{0}^{\Theta_{D}/T} dy \frac{y^4}{e^y-1}.
\label{eqn: memorypho}
\eea
On solving the integral in the above limits, we obtain
\bea
M''_{QQ}(\omega, T)&\propto& T.
\eea
In the case, when $T >> \Theta_{D}$ and $\omega >> T$, the third term of Eq.(\ref{eqn: memoryphononhighfreq}) contributes to the thermal memory function as 
\bea \nonumber
M''_{QQ}(\omega, T) &\approx& \frac{N}{36 \pi^3 m_{i} \rho_{F}^2} \frac{T^7}{\chi_{QQ}^{0}(T)} \left( \frac{q_{D}}{\Theta_{D}} \right)^6   \\ 
&& \times \frac{\omega^2}{ T^2} \int_{0}^{\Theta_{D}/T} dy  \frac{y^4}{e^{y}-1} .
\label{eqn: memoryphononhighomega}
\eea
In the above mentioned frequency and temperature regime, the thermal memory function varies as $\frac{\omega^2}{T}$.\\
In the low temperature limit i.e. $T << \Theta_{D}$, the first term and the third term are the leading order terms in the thermal memory function. Further in the limit $\omega >> T$, 
\bea \nonumber
M''_{QQ}(\omega, T) &\approx& \frac{N}{12 \pi m_{i} \rho_{F}^2} \frac{T^7}{\chi_{QQ}^{0}(T)} \left( \frac{q_{D}}{\Theta_{D}} \right)^6    \\  \nonumber
&& \times \left\lbrace \frac{k_{F}^2}{\pi^2}\left( \frac{\Theta_{D}}{q_{D}T}\right)^2 + \frac{1}{3 \pi^2} \frac{\omega^2}{ T^2}\right\rbrace\int_{0}^{\infty} dy \frac{y^4}{e^{y}-1} . \\
\label{eqn: memoryphononlowomega}
\eea
Similarly in the low frequency limit i.e. when frequency is much smaller than the Debye frequency ($\omega << \omega_{D}$), the Eq.(\ref{eqn: memoryphonon4}) is written as
\bea \nonumber
M''_{QQ}(\omega, T) &\approx& \frac{N}{24 \pi m_{i} \rho_{F}^2} \frac{T^7}{\chi_{QQ}^{0}(T)} \left( \frac{q_{D}}{\Theta_{D}} \right)^6 \frac{\sinh{(\omega/T)}}{\omega/T}  \\ \nonumber
&& \int_{0}^{\Theta_{D}/T} dy \frac{y^5 e^{y}}{(e^y-1)^2} \\
&& \left\lbrace \frac{k_{F}^2}{\pi^2}\left( \frac{\Theta_{D}}{q_{D}T}\right)^2 + \frac{1}{3} - \frac{y^2}{6 \pi^2} \right\rbrace.
\label{eqn: memoryphononlowfreq}
\eea
In the limit $T>> \Theta_{D}$ and $\omega << T$,
\bea \nonumber
M''_{QQ}(\omega, T) &\approx& \frac{N}{36 \pi m_{i} \rho_{F}^2} \frac{T^7}{\chi_{QQ}^{0}(T)} \left( \frac{q_{D}}{\Theta_{D}} \right)^6 \\ \nonumber
&& \times \int_{0}^{\Theta_{D}/T} dy \frac{y^5 e^{y}}{(e^y-1)^2}. \\
\eea
This shows the linear temperature variation and frequency independent character of the thermal scattering rate.\\
In the case when $T << \Theta_{D}$ and $\omega << T$, the Eq.(\ref{eqn: memoryphononlowfreq}) becomes
\bea \nonumber
M''_{QQ}(\omega, T) &\approx& \frac{Nk_{F}^2}{12 \pi^3 m_{i} \rho_{F}^2} \frac{T^5}{\chi_{QQ}^{0}(T)} \left( \frac{q_{D}}{\Theta_{D}} \right)^4 \\
&& \times \int_{0}^{\infty} dy \frac{y^5 e^{y}}{(e^y-1)^2}.
\eea
From the above Eq., we find that $M''_{QQ}(\omega, T)$ varies as $T^{3}$ and frequency independent behavior.\\
In the limit $T << \Theta_{D}$ and $\omega >> T$,
\bea \nonumber
M''_{QQ}(\omega, T) &\approx& \frac{Nk_{F}^2}{24 \pi^3 m_{i} \rho_{F}^2} \frac{T^5}{\chi_{QQ}^{0}(T)} \left( \frac{q_{D}}{\Theta_{D}} \right)^4 \frac{\sinh{(\omega/T)}}{\omega/T} \\
&& \times \int_{0}^{\infty} dy \frac{y^5 e^{y}}{(e^y-1)^2}.
\eea
These analytical predictions of the dynamical behavior of the thermal memory functions in different temperature and frequency domains are supplemented by numerical calculation in the next section. We summarize the above results in the table(\ref{tab:phonontable}). 
\begin{widetext}
\begin{table}
\caption{The thermal scattering rate due to the electron-phonon interaction in different frequency and temperature domains.}
\begin{center}
 \begin{tabular}{|c| c| c| c|} 
 \hline
 $\omega = 0$ & $\omega >> \omega_{D}$ & $\omega << \omega_{D}$ \\ [0.2ex] 
  \begin{tabular}{c | c} 
 \hline
 $T >> \Theta_{D}$ & $T << \Theta_{D}$\\ [0.5ex] 
 \hline
 $1/\tau_{\text{th}} \sim T$ &$1/\tau_{\text{th}} \sim T^{3} $ \\ [0.2ex] 
 \end{tabular} & \begin{tabular}{c| c} 
 \hline
 $\omega >> T$ & $\omega << T$\\ [0.2ex] 
 \hline
 \begin{tabular}{c| c} 
 $T >> \Theta_{D}$ & $T << \Theta_{D}$\\ [0.2ex] 
 \hline
 $1/\tau_{\text{th}} \sim \frac{\omega^2}{T}$ &$1/\tau_{\text{th}} \sim T^3 \left(\frac{k_{F}^2 \Theta_{D}^2}{\pi^2q_{D}^2} + \frac{\omega^2 }{3\pi^2}\right) $ \\ [1ex] 
 \end{tabular} &\begin{tabular}{c} 
  $T >> \Theta_{D}$ \\ [0.2ex] 
 \hline
 $1/\tau_{\text{th}} \sim T$ \\ [1ex] 
 \end{tabular} \\ [1ex] 
\end{tabular} & \begin{tabular}{c| c} 
 \hline
 $\omega >> T$ & $\omega << T$\\ [0.2ex] 
 \hline
 \begin{tabular}{c} 
  $T << \Theta_{D}$ \\ [0.2ex] 
 \hline
 $1/\tau_{\text{th}} \sim T^{4} \frac{\sinh(\omega/T)}{\omega}$ \\ [1ex] 
 \end{tabular} &\begin{tabular}{c| c} 
 $T << \Theta_{D}$ & $T >> \Theta_{D}$\\ [0.2ex] 
 \hline
 $1/\tau_{\text{th}} \sim T^{3}$ &$1/\tau_{\text{th}} \sim T $ \\ [1ex] 
 \end{tabular}\\ [1ex] 
 \end{tabular} \\ [0.2ex] 
 \hline 
  \end{tabular}
  \end{center}
\label{tab:phonontable}
\end{table}
\end{widetext}
\section{Results and Discussion}
\label{sec: results}
In this section, we have plotted the imaginary part of the dynamical thermal memory functions $M''_{QQ}(\omega,T)$ for the case of the electron-impurity and electron-phonon interactions. To extract the characteristic frequency dependent and temperature dependent behavior of $M''_{QQ}(\omega, T)$, we suitably normalize it in various cases. \\

First for the impurity interaction, we plot $M''_{QQ}(\omega,T)/M''_{0}$ where $M''_{0}$ is frequency and temperature independent constant $\left(= \frac{2 k_{F}^4 m}{\pi^5 N_{e}}\right)$, as a function of frequency using the Eq.(\ref{eqn: memoryimpurity3}) in Fig. \ref{fig: impurity_ac_memory}. Here we consider impurity concentration $N_{\text{imp}} = 0.001$ and interaction strength $U=0.1$eV. It is found that the normalized thermal scattering rate increases linearly with the frequency in the range where the frequency is very high as compared to the temperature (as shown in Fig. \ref{fig: ac_impurity_freq_whole}. This linear feature becomes more prominent as the temperature is lowered. For example in Fig. \ref{fig: ac_impurity_freq_whole_small}, the purple curve drawn at $T = 200$K start showing a linear behavior above a frequency lower than that of the other two curves drawn at higher temperatures such as $300$K and $400$K. The low frequency regime $\omega << T$ of the plot is more elaborated in Fig. \ref{fig: ac_impurity_freq_whole_small} which shows deviations from linearity. Also in both the regimes, the thermal scattering rate due to the impurity interaction decreases with the rise in temperature. These features are in accord with our asymptotic analytical predictions (table(\ref{tab:impuritytable})).\\
\figimpurityA

In the zero frequency limit, the thermal scattering rate (Eq. (\ref{eqn: memoryimpurity5})) becomes temperature independent. The same result can be obtained using Boltzmann approach as mentioned in Appendix \ref{app: boltzmann_impurity}. This feature is also in accord with the experimental findings\cite{ziman_book,wilson_book}.\\
\figimpurityE
\figimpurityG
For the electron-phonon interaction, the frequency dependent behavior of the normalized thermal scattering rate (Eq.(\ref{eqn: memoryphonon4})) is shown in Fig. \ref{fig: phonon_ac_memory} at different temperatures. Here the Debye temperature $\Theta_{D}$ is kept fixed at $300$K. In Fig. \ref{fig: ac_phonon_freq_whole}, we observe that in the high frequency regime ($\omega >> \Theta_{D}$),  $M''_{QQ}/M''_{0}$ $\left(M''_{0} = \frac{N m q_{D}^6}{6 \pi^3 m_{i} \rho_{F}^2 N_{e}\Theta_{D}}\right)$ increases as the frequency increases. While in the low frequency regime, it becomes constant . To see the zoomed low frequency behavior, we replot the same curves within a small frequency regime (as shown in Fig. \ref{fig: ac_phonon_freq_whole_small}). We also observe that the magnitude of the thermal memory function reduces with the increase in temperature. However, the exact temperature dependence in the low frequency regime depends on whether the temperature is greater or lower than the Debye temperature. The detail asymptotic behaviors are obtained analytically in previous section (\ref{sec: thermal conductivity}) and given in table(\ref{tab:phonontable}).\\

In Fig. \ref{fig: thermal_ac_phonon_freq}, the real part of the thermal conductivity in case of electron-phonon interaction using Eq.(\ref{eqn: thermal})is plotted as a function of frequency at a fixed Debye temperature $\Theta_{D}$ and at different temperatures. Here we assume that the leading frequency dependence of the thermal conductivity is coming from the thermal scattering rate. Thus to make our discussion simpler, we neglect the frequency dependence of the mass renormalization factor in the thermal conductivity coming from the real part of the thermal memory function. Here we have scaled the frequency with parameter $\omega_{0}$ $\left(= \frac{N m q_{D}^6}{6\pi^3 m_{i}  \rho_{F}^2 N_{e} \Theta_{D}}\right)$, which has the dimension of energy and normalized the real part of the thermal conductivity $\text{Re}[\kappa(\omega, T)]$ with $\kappa_{0}$ $\left(= \frac{\pi^2 N_{e}}{4m \omega_{0}}\right)$. It is observed that the thermal conductivity decays with the increase in frequency in a non-linear manner. Also with the increase of temperature, the thermal conductivity increases. This detail behavior can be understood as follows. Since, our calculation is limited to a perturbative regime i.e. $M''_{QQ}(\omega,T) << \omega$, then $\text{Re}[\kappa(\omega, T)] \sim \frac{\chi_{QQ}^{0}}{T} \frac{M''_{QQ}(\omega, T)}{\omega^2}$. As $\chi_{0}^{QQ}(T) \sim T^2$, thus the real part of the thermal conductivity becomes $\text{Re}[\kappa(\omega, T)] \sim \frac{T M''_{QQ}(\omega, T)}{\omega^2}$. Under this condition, the increase in the thermal conductivity due to the increase in temperature is governed by the factor $T M''_{QQ}(\omega,T)$ which is an increasing function of temperature. Using this relation and table(\ref{tab:phonontable}), various regime of Fig. \ref{fig: thermal_ac_phonon_freq} can be understood. For example, (1) in the regime $T<<\omega<<\omega_{D}$, $\text{Re}[\kappa(\omega,T)] \sim T^5 \frac{\sinh(\omega/T)}{\omega^3}$, (2) in regime $T>>\omega>>\omega_{D}$, $\text{Re}[\kappa(\omega,T)] \sim \frac{T^2}{\omega^2}$, (3) for $\omega >> \omega_{D}$, $\omega >> T$ and $T <<\Theta_{D}$, $\text{Re}[\kappa(\omega,T)] \sim T^4 \left( \frac{a}{\omega^2} + b\right)$, where $a$ and $b$ are constants, etc. The detail asymptotic results of the thermal conductivity due to the electron-phonon and the electron-impurity is given in table(\ref{tab:thermalphonontable}) and (\ref{tab:thermalimpuritytable}). These signatures are new predictions from our formalism and can be verified in future experiments.\\
\begin{widetext}
\begin{table}
\caption{The real part of the thermal conductivity due to the electron-phonon interaction in different frequency and temperature domains.}
\begin{center}
 \begin{tabular}{|c| c| c| c|} 
 \hline
 $\omega = 0$ & $\omega >> \omega_{D}$ & $\omega << \omega_{D}$ \\ 
  \begin{tabular}{c | c} 
 \hline
 $T >> \Theta_{D}$ & $T << \Theta_{D}$\\  
 \hline
 $\kappa \sim T^{0}$ &$\kappa \sim T^{-2} $ \\ 
 \end{tabular} & \begin{tabular}{c| c} 
 \hline
 $\omega >> T$ & $\omega << T$\\  
 \hline
 \begin{tabular}{c| c} 
 $T >> \Theta_{D}$ & $T << \Theta_{D}$\\ 
 \hline
 $\kappa \sim \omega^{0}T^{0}$ &$\kappa \sim T^4\left(\frac{a}{\omega^2} + b\right) $ \\  
 \end{tabular} &\begin{tabular}{c} 
  $T >> \Theta_{D}$ \\ 
 \hline
 $\kappa \sim \frac{T^2}{\omega^2}$ \\ 
 \end{tabular} \\ [1ex] 
\end{tabular} & \begin{tabular}{c| c} 
 \hline
 $\omega >> T$ & $\omega << T$\\ 
 \hline
 \begin{tabular}{c} 
  $T << \Theta_{D}$ \\  
 \hline
 $\kappa \sim T^{5} \frac{\sinh(\omega/T)}{\omega^{3}}$ \\ [1ex] 
 \end{tabular} &\begin{tabular}{c| c} 
 $T << \Theta_{D}$ & $T >> \Theta_{D}$\\ 
 \hline
 $\kappa \sim \frac{T^{4}}{\omega^{2}}$ &$\kappa \sim \frac{T^2}{\omega^2} $ \\ [1ex] 
 \end{tabular}\\ 
 \end{tabular} \\ 
 \hline 
  \end{tabular}
  \end{center}
\label{tab:thermalphonontable}
\end{table}
\end{widetext}

\begin{table}[hbt]
\caption{The real part of the thermal conductivity due to the electron-impurity interaction in different frequency and temperature domains.}
\begin{center}
\begin{adjustbox}{max width = 8.6cm}
 \begin{tabular}{|c| c|} 
 \hline
 $\omega = 0$ & $\omega \ne 0$\\ 
 \hline
 $\kappa \sim T$ & \begin{tabular}{c} 
  \begin{tabular}{c| c} 
 $\omega >> T$ & $\omega << T$\\ 
 \hline
 \begin{tabular}{c} 
  $\kappa \sim \frac{1}{\omega}$ \\ [1ex] 
 \end{tabular} & \begin{tabular}{c} 
 $\kappa \sim \frac{T}{\omega^2} $ \\  
 \end{tabular}
  \end{tabular} \\ 
\end{tabular} \\ [1ex] 
 \hline 
  \end{tabular}
  \end{adjustbox}
  \end{center}
  \label{tab:thermalimpuritytable}
 \end{table}
 
\figimpurityC

Now in the dc limit, we plot $M''_{QQ}(T)/M_{0}$ as a function of temperature $T$ at different Debye's temperatures in Fig. \ref{fig: dc_phonon_temp}. Here we find three important features. One is the increase of the non-linear thermal scattering rate  with temperature in the low temperature regime ($\sim T^3$, refer table(\ref{tab:phonontable})). Second, it increases linearly with the temperature at high temperature regime. Third in the intermediate regime around the Debye temperature, there is a minima in the thermal scattering rate. These features (at high and low temperatures namely $T^3$ at $T<< \Theta_{D}$ and $T$ at $T>>\Theta_{D}$) are in agreement with experiments\cite{ziman_54, rosenberg_55, klemens_56}. In Fig. \ref{fig: dc_thermal_phonon}, using Eq.(\ref{eqn: thermal}) the normalized thermal conductivity has been plotted with temperature $T$. This shows that it decreases as $T^{-2}$  in the low temperature regime and becomes constant in the high temperature regime. These results are consistent with the results derived using Boltzmann approach in Appendix \ref{app: boltzmann_phonon}. In the intermediate temperature regime, it passes through a minimum. This minimum in the thermal conductivity plot is an artifact of neglecting contributions from the Umklapp process in the memory function. Such minima  occurs near the Debye temperature where Umklapp process becomes important. The same peculiarity is also found in Bloch-Boltzmann theory when Umklapp processes are neglected\cite{ziman_book, seitz_book}. 
Such a minima is purely a theoretical artifact and is not observed in any experiments\cite{future}.\\

\section{Conclusion}
\label{sec: conclusion}
Traditionally, the dc transport of a metallic system is discussed in several contexts using Boltzmann equation approach with much success\cite{wilson_book, kasuya_55, navinder_16}. However within this approach, the calculation of the dynamical thermal conductivity is lacking. Also, the Boltzmann approach is solved using relaxation time approximation\cite{nabyendu_16}. On the other hand, the memory function approach is beyond the relaxation time approximation. So, it is a better choice to study the dynamical transport properties in various electronic systems. Also, this approach does not require quasiparticle picture, hence has a broader range of applicability\cite{subir_14,lucas_15, subir_15}. Thus, the memory function formalism is a better choice to study the dynamical transport properties in various electronic systems. However, in the present work, we deal with the system having well defined quasiparticles i.e. metals.

In this work, we perform analytical calculation of the dynamical thermal conductivity of metal for electron-impurity and electron-phonon interactions. We discuss the results in different frequency and temperature domains. Since in the zero frequency limit thermal conductivity of the metal is well known, we consider the results from the Bloch-Boltzmann approach and the experimental findings as a benchmark and compare our results with them.

According to the memory function formalism, the total thermal memory function is the thermal current-thermal current correlation function which captures the role of the impurity and the electron-phonon interactions. This leads the thermal memory function as the sum of the memory functions due to the electron-impurity interactions and the electron-phonon interactions which further result to the total thermal conductivity. We found that at the low temperature, the thermal memory function due to the impurity interaction shows the temperature independent behavior (Eq.(\ref{eqn: memoryimpurity5})). While due to the electron-phonon interaction, it shows $T^3$ behavior (Eq.(\ref{eqn: memoryphonon9})). On the other hand, at the high temperature, the thermal memory function gives linear temperature behavior (Eq.(\ref{eqn: memoryphonon8})).

Now, in the dc limit, the thermal conductivity can be written as
\bea
\kappa(T) &\approx& \frac{T}{M''_{QQ}(T)}, 
\eea
which shows that it varies with an inverse of the memory function. According to the Matthiessen's rule\cite{ziman_book, ashcroft_book}, resistivities add up. Hence, the memory function also add up which is the sum of the memory function due to the electron-impurity and the electron-phonon interactions. Based on that the thermal conductivity can be explained as follows. At very low temperature regime, the conductivity comes mainly due to the impurity interactions which gives the linear temperature dependence behavior. As the temperature increases, the population of the phonon start increasing, resulting the increase of the memory function due to the electron-phonon interaction and the corresponding thermal conductivity decreases. But as the temperature becomes more than the Debye temperature $\Theta_{D}$, the population of the phonon saturates and thus the memory function gives linear temperature dependent behavior and hence the thermal conductivity becomes constant.

In other words, if we consider the impurity and phonon contribution together, we see that the total thermal conductivity can be expressed in an empirical form as,
\bea \nonumber
\frac{1}{\kappa_{\text{total}}(T)} &=& \frac{1}{\kappa_{\text{imp}}(T)} + \frac{1}{\kappa_{\text{ep}}(T)}.\\
&\sim&  \begin{cases}
  \frac{A}{T} + BT^2 ,& \textnormal{at $T<< \Theta_{D}$} \\
 \frac{A}{T} + C ,&\textnormal{at $T>> \Theta_{D}$}.
  \end{cases}
\eea
Here, the first term  and the second term are due to the electron-impurity interaction and the electron-phonon interaction respectively and $A$, $B$ and $C$ are material dependent constants. These results are in accord with the results calculated using Bloch-Boltzmann approach\cite{wilson_book, ziman_book} and also with the experimental findings\cite{ziman_54, rosenberg_55, klemens_56}.

In a general theory of electrical and or thermal conductivity within memory function(matrix) theory must consider the slow relaxation of the conserved total momentum. In principle, one should consider all the relevant slow modes to construct the ``full memory matrix''. The mode with the slowest  relaxation rate is the most relevant in studying the dynamics. Firstly, to keep our discussion simple we neglect the inclusion of the conserved total momentum. However, we see good agreement between our results with that of the previous theories and experiments as well. This is possible because we have confined our discussions on metals with an well defined Fermi surface.

In the finite frequency cases we have several predictions depending on the relative values of the frequency $\omega$, temperature $T$ and the 
Debye frequency $\omega_D$. Few of them can be summarized as follows.
1) $T>>\omega_D$ : in this case, as we move from the low frequency regime to the high frequency regime we see a crossover from the 
$\kappa\sim \frac{T^2}{\omega^2}$ behavior to the $\kappa\sim T^0/\omega^0$ behavior.
2) On the other hand for $T<<\w_D$, we observe that $\kappa\sim \frac{T^4}{w^2}$ in the low frequency regime, then we see  $\kappa\sim T^5 \frac{\sinh{\omega/T}}{\w^3}$  behavior in the intermediate regime and finally see $\kappa\sim T^4 \left(
\frac{a}{\omega^2}+b\right)$ behavior.
These predictions  can be verified in future experiments. Moreover, the present approach can also be used to study other transport properties such as thermo-electric coefficients etc.

\appendix
\section{Thermal conductivity and Memory function relation}
\label{app: thermal_relation_memory}
In the linear response theory, the thermal conductivity is expressed as\cite{kubo_57, zubarev_60, kadanoff_63}
\be
\kappa_{\mu\nu}(z) = \frac{1}{T} \int_{0}^{\infty} dt e^{iz t} \int_{0}^{\beta} d\lambda \langle J_{\nu Q}(-i\hbar\lambda) J_{\mu Q}(t) \rangle.
\ee
Here $\mu$, $\nu$ $=x,y,z$ and represent special directions.

In classical limit i.e. $\hbar \rightarrow 0$, the above Eq. reduces to
\be
\kappa_{\mu\nu}(z) = \frac{1}{T^2} \int_{0}^{\infty} dt e^{iz t} \langle J_{\nu Q}(0) J_{\mu Q}(t) \rangle.
\ee
The time evolution of a dynamical variable $f$ follows Liouville Eq. which is given as
\be
\frac{\partial f}{\partial t} = -\mathcal{L} f,
\ee
where $\mathcal{L}$ is the Liouvillian operator. The solution of the above Eq. yields
\be
f(t) = e^{i\mathcal{L}t}f(0).
\ee
Using the above relation, the Kubo formula for the thermal conductivity can be written as
\be
\kappa_{\mu\nu}(z) = \frac{1}{T^2} \int_{0}^{\infty} dt e^{izt} \langle J_{\nu Q}(0) e^{i\mathcal{L}t} J_{\mu Q}(0) \rangle.
\ee
On further simplification, it becomes
\be
\kappa_{\mu\nu}(z) = \frac{1}{T^2} \left\langle J_{\nu Q} \left\vert \frac{i}{z+\mathcal{L}} \right\vert J_{\mu Q} \right\rangle.
\label{eqn: kmunu}
\ee
Now we introduce the projection operator $\mathcal{P}$ which is defined as follows 
\bea
\mathcal{P} &=& \sum_{\nu, \mu} \frac{\vert J_{\nu Q} \rangle \langle J_{\mu Q} \vert}{\langle J_{\nu Q} \vert J_{\mu Q} \rangle} = \mathcal{I} - \mathcal{Q},
\eea
where $\mathcal{I}$ is an identity matrix and $Q= \mathcal{I}  - \mathcal{P}$ is an unprojected part. Then replace $\mathcal{L}$ by $\mathcal{L}(\mathcal{P}+\mathcal{Q})$ in Eq.(\ref{eqn: kmunu}),$\kappa_{\mu\nu}(z)$ becomes
\bea \nonumber
\kappa_{\mu\nu}(z) &=& i\frac{1}{T^2} \left\langle J_{\nu Q} \left\vert \frac{i}{z+\mathcal{L} Q} \right\vert J_{\mu Q} \right\rangle \\
&&- i\frac{1}{T^2} \left\langle J_{\nu Q} \left\vert \frac{i}{z+\mathcal{L} Q} \mathcal{L} \mathcal{P} \frac{1}{z+\mathcal{L}} \right\vert J_{\mu Q} \right\rangle.
\label{eqn: therm}
\eea
On expanding the above Eq., the first term is $i\frac{1}{zT^2} \langle J_{\nu Q}\vert J_{\mu Q} \rangle$ which can be written as $i\frac{\chi_{QQ}^{0}(T)}{T z}$ where $\chi_{QQ}^{0}(T)$ is the static thermal current-thermal current correlation function. Inserting the projection operator into the second term, the later becomes
\be
\frac{1}{T^2} \left\langle J_{\nu Q} \left\vert \frac{i}{z+\mathcal{L} Q} \mathcal{L} \sum_{\mu ' Q} \vert J_{\mu ' Q} \rangle \langle J_{\mu ' Q} \vert \frac{1}{z+\mathcal{L}} \right\vert J_{\mu Q} \right\rangle.
\ee
Inserting the above expressions of the first and the second term in Eq.(\ref{eqn: therm}), the thermal conductivity in the isotropic case can be written as
\be
\kappa(z,T) = i \frac{1}{T} \frac{\chi_{QQ}^{0}(T)}{z+ M_{QQ}(z, T)},
\ee
where $M_{QQ}(z,T)$ is the thermal memory function
\be
M_{QQ}(z, T) = \frac{1}{T \chi_{QQ}^{0}(T)} \left\langle J_{Q} \left\vert \frac{z}{z+ \mathcal{L}\mathcal{Q}} \mathcal{L} \right\vert J_{Q} \right\rangle.
\ee
\section{Derivation of static correlation function}
\label{app: static_derivation}
The static thermal current-thermal current correlation is defined as\cite{chernyshev_15}
\be
\chi_{QQ}^{0}(T) = \frac{1}{3T} \sum_{\textbf{k}} (\epsilon_{\textbf{k}} - \mu)^2 v_{\textbf{k}}^2 f_{\textbf{k}} (1-f_{\textbf{k}}).
\ee
Converting the summation into energy integral and substituting $\frac{\epsilon_{\textbf{k}} - \mu}{T} = \eta$, the above Eq. reduces to
\bea \nonumber
\chi_{QQ}^{0}(T) &=& \frac{T^2 k_{F}^3}{3m} \frac{1}{2\pi^2} \int_{0}^{\infty} d\eta \frac{\eta^2 e^{\eta}}{(e^{\eta}+1)^2}. \\
&=& T^2  \frac{N_{e}}{m} \frac{\pi^2}{12}
\label{eqn: static_expression}
\eea
This shows that the static thermal current-thermal current correlation varies quadratically in temperature.
\section{Thermal conductivity using Boltzmann approach}
\subsection{For Impurity interaction}
\label{app: boltzmann_impurity}
The Boltzmann equation for the semiclassical distribution function $g_k(\textbf{r},t)$ is written as
\be
v_{\textbf{k}} \frac{\partial g_k}{\partial r} = \left(\frac{\partial g_k}{\partial t} \right)_{\text{coll}} = \int \frac{d\textbf{k}'}{2\pi^3} (W(\textbf{k}' \rightarrow \textbf{k})-W(\textbf{k}\rightarrow \textbf{k}')).
\label{eqn: collisionterm}
\ee
Here $W(\textbf{k}' \rightarrow \textbf{k})$ defines the transition probability of an electron scattering from initial state $\textbf{k}'$ to final state $\textbf{k}$. According to the Fermi-Golden rule, in case of the impurity scattering it can be expressed as
\bea \nonumber
W(\textbf{k}' \rightarrow \textbf{k}) &=& 2\pi \vert \langle \textbf{k}' \vert H_{\text{imp}} \vert \textbf{k} \rangle \vert^2 \delta(\epsilon_{\textbf{k}'}-\epsilon_{\textbf{k}}).\\
\eea
Considering the impurity interaction Hamiltonian given in Eq.(\ref{eqn: impurityhamiltonian}), the transition probability can be expressed as
\bea \nonumber
W(\textbf{k}' \rightarrow \textbf{k}) &=& 4\pi \frac{N_\text{imp}}{N^2} \vert U(\textbf{k}', \textbf{k}) \vert^2 g_{\textbf{k}} (1-g_{\textbf{k}'}) \delta(\epsilon_{\textbf{k}'}-\epsilon_{\textbf{k}}).\\
\eea
Here $U(\textbf{k}', \textbf{k}) = \langle \textbf{k}' \vert U \vert \textbf{k} \rangle $, the matrix element for the impurity interaction. Inserting the above Eq. in Eq.(\ref{eqn: collisionterm}), we obtain
\bea \nonumber
\left(\frac{\partial g_k}{\partial t} \right)_{\text{coll}} &=& \int d\textbf{k}' \frac{N_\text{imp}}{2\pi^2N^2} \vert U(\textbf{k}', \textbf{k}) \vert^2 \left(g_{\textbf{k}'} -g_{\textbf{k}}  \right)\delta(\epsilon_{\textbf{k}'}-\epsilon_{\textbf{k}}).\\
\label{eqn: collisionwithUimpurity}
\eea
Now linearizing the Boltzmann equation using $g_{\textbf{k}} = f_{\textbf{k}} + \delta g_{\textbf{k}}$ and taking equilibrium collision integral terms to zero, the Eq.(\ref{eqn: collisionwithUimpurity}) can be written as
\bea \nonumber
\left(\frac{\partial g_k}{\partial t} \right)_{\text{coll}} &=& \int d\textbf{k}' \frac{N_\text{imp}}{2\pi^2N^2} \vert U(\textbf{k}', \textbf{k}) \vert^2 \left(\delta g_{\textbf{k}'} -\delta g_{\textbf{k}}  \right)\delta(\epsilon_{\textbf{k}'}-\epsilon_{\textbf{k}}).\\
\label{eqn: collisionwithUimpurity2}
\eea
In the standard procedure, the collision integral is solved by an iterative procedure\cite{wilson_book, kasuya_55, navinder_16}. One starts with the relaxation time approximation.
\be
g_{k} = f_{k} + \delta g_{k} = f_{k} + \frac{k_{x}}{m} \tau(\epsilon_{\textbf{k}}) \left(\frac{\partial f_{k}}{\partial T}\right) (\nabla T)_{x}.
\ee
Thus the change in the distribution function is written as
\be
\delta g_{k} = g_{k}-f_{k} =  \frac{k_{x}}{m} C(\epsilon_{k}) \left(\frac{\partial f_{k}}{\partial \epsilon}\right),
\ee
Here $C(\epsilon_{k})$ is proportional to an energy dependent relaxation time. On substituting the above expression in Eq.(\ref{eqn: collisionwithUimpurity2}) and noticing that $v_{\textbf{k}}^{x} \nabla g_{\textbf{k}} = \frac{k_{x}}{m} \frac{\partial f_{k}}{\partial T} \nabla T $, one obtains
\bea \nonumber
\frac{1}{\tau(\epsilon_{\textbf{k}})} &=& \frac{2 N_\text{imp} m k_{F}}{\pi N^2} \int_{0}^{\pi} d\theta \vert U(k_{F},\theta) \vert^2 \sin\theta (1-\textbf{k}.\textbf{k}').\\
\label{eqn: boltzmann_impurity_scattering}
\eea
This shows that the thermal scattering rate due to impurity interaction is independent of the temperature. As the thermal conductivity is defined as
\bea \nonumber
\kappa(T) &=& \frac{2}{T^2} \sum_{\textbf{k}} \tau(\epsilon_{\textbf{k}})\left(\epsilon_{\textbf{k}} - \mu \right)^{2}\frac{e^{(\epsilon_{\textbf{k}}-\mu)/T}}{\left( e^{(\epsilon_{\textbf{k}}-\mu)/T} +1 \right)^2}.\\  
\eea
Substituting the Eq.(\ref{eqn: boltzmann_impurity_scattering}) in the above Eq., the thermal conductivity due to the electron-impurity interaction shows the temperature dependence as
\bea \nonumber
\kappa(T) &=& \frac{1}{72} \frac{\pi k_{F}^{2}}{N_{\text{imp}} U^{2}m^{2}} T\\
\text{i.e.} \hspace*{2mm}\kappa(T) &\propto& T.
\label{eqn: thermal_boltzmann_impurity}
\eea
From this we infer that the results of the thermal conductivity using both the approaches the memory function and the Boltzmann approach agree quantitatively to each other.
\subsection{For electron-phonon interaction}
\label{app: boltzmann_phonon}
Similarly for the electron-phonon interaction case, the Boltzmann equation becomes
\be
v_{\textbf{k}} \frac{\partial g_k}{\partial r} = \left(\frac{\partial g_k}{\partial t} \right)_{\text{coll}} = \int d\textbf{k} (W(\textbf{k}+\textbf{q} \rightarrow \textbf{k})-W(\textbf{k}\rightarrow \textbf{k}+\textbf{q})).
\ee
Here $W(i\rightarrow f)$ is the transition probability involving both the emission and absorption of phonons. This, using Fermi Golden rule can be expressed as\cite{ashcroft_book}
\bea \nonumber
W(\textbf{k}+ \textbf{q} \rightarrow \textbf{k}) &=& 2\pi \vert \langle \textbf{k} \vert H_{\text{ep}} \vert \textbf{k}+\textbf{q} \rangle \vert^2 \delta(\epsilon_{\textbf{k}+\textbf{q}}-\epsilon_{\textbf{k}}\pm \omega_{q}).\\
\eea
Using the Eq.(\ref{phononhamiltonian}), above expression for the transition probability can be written as
\bea \nonumber
W(\textbf{k}+ \textbf{q} \rightarrow \textbf{k}) &=& 4\pi \vert D(q) \vert^2 g_{\textbf{k}+\textbf{q}}(1-g_{\textbf{k}})(n_{\textbf{q}}+ 1)\\
&& \delta(\epsilon_{\textbf{k}} + \omega_{q} - \epsilon_{\textbf{k}+\textbf{q}}).
\eea
Considering all possible scattering processes, the collision integral can be written as
\bea \nonumber
\left( \frac{\partial g_{k}}{\partial t} \right)_{\text{coll}} &=& \int d\textbf{q} \left( U(\textbf{k}+\textbf{q}:\textbf{k}) g_{\textbf{k}+\textbf{q}} (1-g_{\textbf{k}}) \right. \\
&& \left.- U(\textbf{k};\textbf{k}+\textbf{q})g_{\textbf{k}}(1-g_{\textbf{k}+\textbf{q}}) \right),
\label{eqn: collisionintegral}
\eea
where
\bea \nonumber
U(\textbf{k}+\textbf{q};\textbf{k}) &=& W_{\textbf{q}}^{0} [(n_{\textbf{q}}+1)\delta(\epsilon_{\textbf{k}} + \omega_{q} - \epsilon_{\textbf{k}+\textbf{q}}) \\
&&+ n_{-\textbf{q}}\delta(\epsilon_{\textbf{k}} - \omega_{q} - \epsilon_{\textbf{k}+\textbf{q}})]
\eea
\bea \nonumber
U(\textbf{k};\textbf{k}+\textbf{q}) &=& W_{\textbf{q}}^{0} [(n_{\textbf{-q}}+1)\delta(\epsilon_{\textbf{k}+\textbf{q}} + \omega_{q} - \epsilon_{\textbf{k}}) \\
&&+ n_{\textbf{q}}\delta(\epsilon_{\textbf{k}+\textbf{q}} - \omega_{q} - \epsilon_{\textbf{k}})],
\eea
and $W_{\textbf{q}}^{0} = 4\pi \vert D(\textbf{q}) \vert^2$.\\
The details of the calculation is given in the references (\cite{wilson_book, kasuya_55, navinder_16}). Here we note that using the relation $U(\textbf{k}+\textbf{q};\textbf{k}) = e^{\beta\epsilon_{\textbf{k}+\textbf{q}}} e^{-\beta\epsilon_{\textbf{k}}} U(\textbf{k};\textbf{k}+\textbf{q})$ and linearizing the Boltzmann equation by substituting $g_{k} = f_{k} + \delta g_{k}$ and taking the equilibrium collision integral terms to be zero, the Eq.(\ref{eqn: collisionintegral}) can be reduced to,
\bea \nonumber
\left( \frac{\partial g_{k}}{\partial t} \right)_{\text{coll}} &=& \int d\textbf{q} U(\textbf{k};\textbf{k}+\textbf{q})\left\lbrace \delta g_{\textbf{k}+\textbf{q}}(e^{-\beta(\epsilon_{\textbf{k}}-\epsilon_{\textbf{k}+\textbf{q}})} \right. \\ \nonumber
&&\left. (1-f_{\textbf{k}}) + f_{\textbf{k}}) - \delta g_{\textbf{k}} (e^{-\beta(\epsilon_{\textbf{k}}-\epsilon_{\textbf{k}+\textbf{q}})}f_{\textbf{k}+\textbf{q}}\right. \\
&& \left. + (1-f_{\textbf{k}+\textbf{q}})\right\rbrace.
\eea
On further simplifications, the collision integral can be written as
\bea \nonumber
\left( \frac{\partial g_{k}}{\partial t}\right)_{\text{coll}}  &=& \beta \int d\textbf{q} W_{\textbf{q}}^{0} n_{\textbf{q}} \left( f_{\textbf{k}+\textbf{q}}(1-f_{\textbf{k}}) \delta(\epsilon_{\textbf{k}+\textbf{q}} + \omega_{-q} - \epsilon_{\textbf{k}})\right. \\ \nonumber
&& \left. + f_{\textbf{k}}(1-f_{\textbf{k}+\textbf{q}}) \delta(\epsilon_{\textbf{k}+\textbf{q}} - \omega_{q} - \epsilon_{\textbf{k}}\right)\\
&& (\delta\phi(\textbf{k}+\textbf{q}) - \delta\phi(\textbf{k})),
\label{eqn: coll}
\eea
where $\delta \phi(\textbf{k}) = \frac{\delta g_{\textbf{k}}}{\beta f_{\textbf{k}}(1-f_{\textbf{k}})}$.\\
As explained in the impurity scattering case that the calculation is done by an iterative procedure, where one introduces
\be
\delta \phi(k) = \frac{k_{x}}{m} C(\epsilon_{k}).
\label{eqn: phidefinition}
\ee
From Eq.s (\ref{eqn: coll}) and (\ref{eqn: phidefinition}), we have
\bea \nonumber
\frac{k_{x}}{m}\left( \frac{\partial f_{k}}{\partial T} \right) (\nabla T)_{x} &=& \left( \frac{\partial g_{k}}{\partial t} \right)_{\text{coll}} \\ \nonumber
&=& \frac{4\pi}{m T} \int d\textbf{q} \vert D(\textbf{q})\vert^2 n_{\textbf{q}} \\ \nonumber
&&\left\lbrace f_{\textbf{k}+\textbf{q}}(1-f_{\textbf{k}}) \delta(\epsilon_{\textbf{k}+\textbf{q}} + \omega_{-q} - \epsilon_{\textbf{k}})\right. \\ \nonumber
&& \left. + f_{\textbf{k}}(1-f_{\textbf{k}+\textbf{q}}) \delta(\epsilon_{\textbf{k}+\textbf{q}} - \omega_{q} - \epsilon_{\textbf{k}})\right\rbrace \\ \nonumber
&& \left\lbrace (k_{x}+q_{x}) C(\epsilon_{\textbf{k}+\textbf{q}}) - k_{x}  C(\epsilon_{\textbf{k}})\right\rbrace. \\
\eea
On inserting the phonon matrix element, solving the angular integrals and introducing the dimensionless variables $\frac{\epsilon_{\textbf{k}} - \mu}{T} = \eta$ and $\frac{\omega_{q}}{T} = z$, the collision integral reduces to
\bea \nonumber
\left( \frac{\partial g_{k}}{\partial t} \right)_{\text{coll}} &=& -\frac{1}{2\pi m_{i} N \rho_{F}^2 (2m)^{1/2}} \epsilon^{-3/2} k_{x} \frac{\partial f_{k}}{\partial\epsilon} \left( \frac{T}{\Theta_{D}} \right)^3 \frac{q_{D}^4}{\Theta_{D}}\\ \nonumber
&& \int_{0}^{\Theta_{D}/T} dz  \frac{z^2}{e^z -1} \\ \nonumber
&& \left\lbrace  \frac{e^{\eta} + 1}{e^{\eta -z}+1} \left[\left( \epsilon - \frac{1}{2} D \left(\frac{T}{\Theta_{D}}\right)^2 z^2 - \frac{1}{2} Tz \right) \right.\right. \\ \nonumber
&&\left.\left.  C(\eta - z) -\epsilon  C(\eta) \right] + \frac{e^{z}(e^{\eta} + 1)}{e^{\eta +z}+1} \right. \\ \nonumber
&& \left.  \left[\left( \epsilon - \frac{1}{2} D \left(\frac{T}{\Theta_{D}}\right)^2 z^2 + \frac{1}{2} Tz \right) \right.\right.\\
&& \left.\left.  C(\eta + z) -\epsilon  C(\eta) \right] \right\rbrace.
\eea
Here $D = \frac{q_{D}^{2}}{2m}$. On further simplifications, the above expression can be written as
\bea \nonumber
-\frac{k_{x}}{m} \eta \left( \frac{\partial f_{k}}{\partial \epsilon} \right) (\nabla T)_{x} &=& \left( \frac{\partial g_{k}}{\partial t} \right)_{\text{coll}}\\ \nonumber
&=& -\frac{k_{x}}{2\pi m_{i} N \rho_{F}^2} \frac{\epsilon^{-3/2}}{(2m)^{1/2}}  \frac{\partial f_{k}}{\partial \epsilon} \left( \frac{T}{\Theta_{D}} \right)^3 \frac{q_{D}^4}{\Theta_{D}} \\ \nonumber
&& \times \int_{-\Theta_{D}/T}^{\Theta_{D}/T} dz \frac{z^2 }{\vert e^z -1 \vert} \frac{e^{\eta} + 1}{e^{\eta +z}+1} \\ \nonumber
&&  \left[\left( \epsilon - \frac{1}{2} D \left(\frac{T}{\Theta_{D}}\right)^2 z^2 + \frac{1}{2} Tz \right) \right. \\ \nonumber
&&\left.  C(\eta + z) -\epsilon  C(\eta) \right]. \\
\eea
In the above Eq., the contribution from the terms with odd power in $z$ vanishes. Thus on simplification, we have
\bea \nonumber
&&\frac{2 \pi m_{i} N \rho_{F}^2\epsilon_{F}^{1/2} (2m)^{1/2}}{m} \frac{\Theta_{D}}{q_{D}^4}\left( \frac{\Theta_{D}}{T} \right)^3  \eta  (\nabla T)_{x}\\ \nonumber
&=&  \int_{-\Theta_{D}/T}^{\Theta_{D}/T} dz \frac{z^2 }{\vert e^z -1 \vert} \frac{e^{\eta} + 1}{e^{\eta +z}+1} \\ \nonumber
&&  \left[\left( 1 - \frac{ D}{2\epsilon_{F}} \left(\frac{T}{\Theta_{D}}\right)^2 z^2 \right) C(\eta + z) -  C(\eta) \right]. \\
\label{eqn: simplifyiedexpression}
\eea
In the high temperature limit i.e. $T >> \Theta_{D}$, the term within the bracket in Eq.(\ref{eqn: simplifyiedexpression}) with $T^2$ contributes more then the others terms and in the case $\eta >> z$, the $C(\eta)$ can be approximated as
\bea \nonumber
C(\eta) &\approx& -\frac{16 \pi m_{i} \rho_{F}^2 N \epsilon_{F}^{3/2} (2m)^{1/2} \Theta_{D}}{m D q_{D}^4} \left( \frac{\Theta_{D}}{T} \right) \eta (\nabla T)_{x}.
\eea
The thermal current is defined as
\bea \nonumber
J_{Q} &=& 2 \int \frac{d\textbf{k}}{(2\pi)^3} v_{\textbf{k}} (\epsilon_{\textbf{k}} -\mu) \delta g_{\textbf{k}} \\
&=& \frac{2k_{F}^3}{\pi^2} \int d\eta \eta  C(\eta) \frac{\partial f_{k}}{\partial \eta} .
\label{eqn: thermaldefi}
\eea
Substituting the value of $C(\eta)$ and using the relation $J_{Q} = -\kappa (\nabla T)_{x}$, we find that the thermal conductivity in high temperature regime becomes
\bea \nonumber
\kappa(T) &\approx& \frac{8}{3} \frac{\pi k_{F}^{6} m_{i} \rho_{F}^{2} \Theta_{D}^{2} N}{q_{D}^{6} m^{2}} \\
\text{i.e.} \hspace{2mm} \kappa(T) &=& \text{constant}.
\eea
Now in the case of low temperature ($T << \Theta_{D}$), the right hand side of Eq.(\ref{eqn: simplifyiedexpression}) can be written as
\bea \nonumber
\int_{-\Theta_{D}/T}^{\Theta_{D}/T} dz \frac{z^2 }{\vert e^z -1 \vert} \frac{e^{\eta} + 1}{e^{\eta +z}+1}  \left[ C(\eta + z) -  C(\eta) \right].\\
\label{eqn: lowtemperatureappendix}
\eea
The above Eq. can be solved by variational method\cite{kasuya_55}. Following the reference (\cite{kasuya_55}), in the low temperature limit, we can write,
\bea \nonumber
C(\eta) &=& -\frac{4\pi \Theta_{D} \epsilon_{F}^{1/2} \rho_{F}^2 m_{i} N}{3 m q_{D}^4} \left( \frac{\Theta_{D}}{T} \right)^3 \eta (\nabla T)_{x}. \\
\eea
Substituting the above Eq. in (\ref{eqn: thermaldefi}), we observe that the thermal conductivity shows a temperature dependence of the following form
\bea \nonumber
\kappa(T) &\approx& \frac{2}{125} \frac{\pi^{3} k_{F}^{4}  m_{i}\rho_{F}^{2} \Theta_{D}^{4} N}{m^{2} q_{D}^{4}} \\
\kappa(T) &\propto& T^{-2}.
\eea
Thus, we see that the thermal conductivity in the case of electron-phonon interaction shows inverse square temperature dependence in the low temperature regime and saturates to a constant value in the high temperature regime within the Bloch-Boltzmann approach and this agrees qualitatively with our calculation using the memory function formalism. Because of the approximate results of the thermal conductivity, the numeric factors are different in the thermal conductivity expressions in both the approaches.


\begin{thebibliography}{}

\bibitem{bridges_02}
D. Cao, F. Bridges, G. R. Kowach and A. P. Ramirez, Phys. Rev. Lett., \textbf{89},  215902 (2002).

\bibitem{bhalla_14} 
P. Bhalla and N. Singh, Eur. Phys. J. B, \textbf{87}, 213 (2014).

\bibitem{chernyshev_15}
A. L. Chernyshev and W. Brenig, Phys. Rev. B, \textbf{92}, 054409 (2015).

\bibitem{jain_16}
A. Jain and A. J. H. McGaughey, Phys. Rev. B, \textbf{93}, 081206 (2016).

\bibitem{romano_16}
G. Romano, K. Esfarjani, D. A. Strubbe, D. Broido and A. M. Kolpak, Phys. Rev. B, \textbf{93}, 035408 (2016).

\bibitem{bidwell_40}
C. C. Bidwell, Phys. Rev. \textbf{58}, 561 (1940).

\bibitem{deo_65}
B. Deo and S. N. Behera, Phys. Rev. \textbf{141}, 738 (1966).

\bibitem{wilson_book}
A. H. Wilson, The Theory of Metals (Cambridge University Press, 1953).

\bibitem{ziman_book} 
J. M . Ziman, Electrons and Phonons (Clarendon Oxford, 1960).

\bibitem{volz_01}
S. G. Volz, Phys. Rev. Lett., \textbf{87}, 074301 (2001).

\bibitem{koh_07}
Y. K. Koh and D. G. Cahill, Phys. Rev. B, \textbf{76}, 075207 (2007).

\bibitem{shastry_06}
B. S. Shastry, Phys. Rev. B, \textbf{73}, 085117 (2006).

\bibitem{shastry_09}
B. S. Shastry, Rep. Prog. Phys., \textbf{72}, 016501 (2009).

\bibitem{dhar_11}
A. Dhar, O. Narayan, A. Kundu and K. Saito, Phys. Rev. E, \textbf{83}, 011101 (2011).

\bibitem{ezzahri_12}
Y. Ezzahri and K. Joulain, J. Appl. Phys., \textbf{112}, 083515 (2012).

\bibitem{yang_15}
F. Yang and C. Dames, Phys. Rev. B, \textbf{91}, 165311 (2015).

\bibitem{zwanzig_61a}
R. Zwanzig, Phys. Rev. \textbf{124}, 983 (1961).

\bibitem{zwanzig_61} 
R. Zwanzig, in Lectures in Theoretical Physics, edited by W. E. Brittin, B. W. Downs and J. Downs (Interscience, New York, 1961), vol. 3, p. 135.

\bibitem{mori_65}
H. Mori, Progr. Theoret. Phys. \textbf{33}, 423 (1965).

\bibitem{forster_95}
D. Forster, Hydrodynamic Fluctuations, Broken Symmetry And Correlation Functions, (Advanced Books Classics 1995).

\bibitem{fulde_12}
P. Fulde, Correlated electrons in Quantum Matter (World Scientific, 2012).

\bibitem{pires_88}
A. S. T. Pires, Helvetica Physica Acta, \textbf{61}, 988 (1988).

\bibitem{berne_66}
B. J. Berne, J. P. Boon and S. A. Rice, J. Chem. Phys. \textbf{45}, 1086 (1966).

\bibitem{harp_70}
G. D. Harp and B. J. Berne, Phys. Rev. A, \textbf{2}, 975 (1970).

\bibitem{berne_70}
B. J. Berne and G. D. Harp, Advan. Chem. Phys. \textbf{XVII}, 63 (1970).

\bibitem{arfi_92} 
B. Arfi, Phys. Rev. B, \textbf{45}, 2352 (1992).

\bibitem{maldague_77}
P. F. Maldague, Phys. Rev. B, \textbf{16}, 2437 (1977).

\bibitem{nabyendu_15}
N. Das and N. Singh, IJMP B, \textbf{30}, 1650071 (2016).

\bibitem{nabyendu_16}
N. Das, P. Bhalla and N. Singh, ArXiv e-prints, arXiv:1601.01127 (2016).

\bibitem{kadanoff_63}
L. P. Kadanoff and P. C. Martin, Ann. Phys. \textbf{24}, 419 (1963).

\bibitem{kubo_57} 
R. Kubo, J. Phys. Soc. Japan \textbf{12}, 570 (1957).

\bibitem{zubarev_60}
D. N. Zubarev, Usp. Fiz. Nauk \textbf{71}, 71 (1960).

\bibitem{gotze_72} 
W. G\"otze and P. W\"olfle, Phys. Rev. B, \textbf{6}, 1226 (1972).

\bibitem{navinder_16}
N. Singh, Electronic Transport Theories: From Weakly to Strongly Correlated Materials, (Taylor and Francis Group, CRC Press, 2016).

\bibitem{bhalla_16}
P. Bhalla and N. Singh, Eur. Phys. J. B, \textbf{89}, 49 (2016).

\bibitem{bhalla_16a}
P. Bhalla, N. Das and N. Singh, Phys. Lett. A, \textbf{380}, 2000 (2016).

\bibitem{mahan_book}
G. D. Mahan, Many-Particle Physics (Plenum, New York and London, 2nd. Ed.,1990).

\bibitem{ziman_54}
J. M. Ziman, Proc. Roy. Soc. A, \textbf{226}, 436 (1954).

\bibitem{rosenberg_55}
H. M. Rosenberg, Phil. Trans. Roy. Soc. A, \textbf{247}, 441 (1955).

\bibitem{klemens_56}
P. G. Klemens, Thermal Conductivity of Solids at Low Temperatures, \textbf{14} (Springer Verlag, Berlin, 1956). 

\bibitem{seitz_book}
F. Seitz and D. Turnbull, Solid State Physics, Advances in Research and Applications, \textbf{4} (Academic Press Inc. Publishers, New York, 1957).

\bibitem{future}
A calculation based on the memory function formalism including both N-process and U-processs is planned for a future investigation.

\bibitem{ashcroft_book}
 N.W. Ashcroft, N.D. Mermin, \textit{Solid state physics}, Science: Physics (Saunders College, 1976). 

\bibitem{kasuya_55} 
T. Kasuya, Prog. Theor. Phys., \textbf{13}, 561 (1955).


\bibitem{subir_14}
A. A. Patel and S. Sachdev, Phys. Rev. B \textbf{90}, 165146 (2014).

\bibitem{lucas_15}
A. Lucas, Journal of High Energy Physics, \textbf{2015}, 1  (2015).

\bibitem{subir_15}
A. Lucas and S. Sachdev, Phys. Rev. B \textbf{91}, 195122 (2015).







\end{thebibliography}
\end{document}